\documentclass[amsmath, twocolumn, tighten, times]{aastex702}

\usepackage{colortbl}
\usepackage[T1]{fontenc}

\usepackage[normalem]{ulem}
\usepackage{orcidlink}

\begin{document}

\title{Beyond static torques: Self-consistent migration and dynamical trapping of low-mass planets in dusty disks}

\author[orcid=0009-0009-4329-2916,sname='Gomez', gname=Jose L.]{Jose L. Gomez}
\affiliation{Facultad de Ciencias Astronomicas y Geofisicas, Universidad Nacional de La Plata, Paseo del Bosque s/n, 1900 La Plata, Argentina}
\affiliation{Instituto de Astrofísica de La Plata (IALP), CCT La Plata-CONICET-UNLP, Paseo del Bosque s/n, La Plata, Argentina}
\email[show]{jlgomez@fcaglp.unlp.edu.ar}

\author[orcid=0000-0002-3728-3329,gname=Pablo, sname=Benítez-Llambay]{Pablo Benítez-Llambay} 
\affiliation{Facultad de Ingenier{\'\i}a y Ciencias Universidad Adolfo Ib\'a\~nez, Av. Diagonal las Torres 2640, Pe\~nalol\'en, Chile}
\email[show]{pablo.benitez@uai.cl}

\author[orcid=0000-0002-2450-7389,gname=Ximena S.,sname=Ramos]{Ximena S. Ramos}
\affiliation{Centro Multidisciplinario de Física, Vicerrectoría de Investigación, Universidad Mayor, 8580745 Santiago, Chile}
\email[]{ximena.ramos@umayor.cl}

\author[orcid=0000-0001-8577-9532,gname=Octavio M.,sname=Guilera]{Octavio M. Guilera}
\affiliation{Facultad de Ciencias Astronomicas y Geofisicas, Universidad Nacional de La Plata, Paseo del Bosque s/n, 1900 La Plata, Argentina}
\affiliation{Instituto de Astrofísica de La Plata (IALP), CCT La Plata-CONICET-UNLP, Paseo del Bosque s/n, La Plata, Argentina}
\email[]{oguilera@fcaglp.unlp.edu.ar}

\begin{abstract}
The rapid inward migration of low-mass planets remains a primary theoretical challenge in planet formation. While recent studies show that the solid component of protoplanetary disks can exert strong positive torques to counteract the negative gas torques, traditional static fixed-orbit models fail to capture the fully coupled dynamics of a migrating planet. In this work, we present 2D high-resolution, two-fluid hydrodynamical simulations using a comoving reference frame to track the self-consistent orbital evolution of embedded planets across a comprehensive six-dimensional parameter space. By allowing the planet to dynamically respond to disk-exerted torques, we identify a wide diversity of migration regimes, involving inward/outward migration and oscillatory radial confinement. We also find that massive cores can overcome dust-induced barriers, experiencing dust-mediated runaway outward or inward migration. Conversely, in dust-rich environments continuous gas-dust momentum exchange combined with planet torques dynamically regenerates co-orbital dust asymmetries, creating robust dynamical attractors where super-Earths undergo stable, oscillatory confinement, effectively halting inward drift without requiring pre-existing pressure bumps or disk substructures. This self-induced trapping mechanism could be relevant in the outer disk, beyond the water-ice line in environments with a enhanced local dust-to-gas ratio, where icy pebbles are moderately coupled to the gas. In low-metallicity or highly diffusive disks, planets generally adopt monotonic inward decay, apart from cases with outward migration. Ultimately, the diversity of these self-consistent trajectories demonstrates that static torque approximations are insufficient. Evaluating fully coupled kinematic feedback is essential for accurately predicting planetary survival and the final architecture of planetary systems. 
\end{abstract}


\section{Introduction}
\label{sec:introduction}

Planets form, grow, and evolve embedded in the gaseous and dusty environments of protoplanetary disks. The gravitational field of a forming embryo perturbs the surrounding disk material, generating asymmetric structures such as spiral density wakes. These structures, in turn, exert a net gravitational torque on the planet, causing its orbit to evolve and change. This process, known as planetary migration, is a fundamental component of modern planet formation theory, as its characterization is crucial to the development of predictive models that can explain the observed architecture of exoplanetary systems \citep[][]{Paardekooper2023, Benitez-Llambay2026}.

For low-mass planets, theoretical models face a significant challenge known as rapid inward migration. In this so-called type-I regime, the orbital decay timescale is often significantly shorter than the million-year lifetime of the protoplanetary disk \citep{Tanaka2002}. This rapid orbital decay makes it difficult to explain the survival of the vast population of observed super-Earths and the formation of giant planet cores, which must remain in favorable disk regions to accrete gas.

In the classical gas-driven migration framework, the total torque exerted on a planet is determined by a delicate balance between the differential Lindblad torque, which is typically negative and drives inward migration, and the corotation torque, which can be positive but is highly sensitive to the disk's local physical properties \citep[e.g.][]{Paardekooper2011, JimenezMasset2017}. While more sophisticated models including thermal effects \citep[e.g][]{Benitez-Llambay2015, Masset2017} have identified scenarios where migration can be slowed, halted, or reversed, a universal mechanism for preventing the rapid inward drift of low-mass planets remains elusive.

\begin{figure*}[th!]
    \centering
\includegraphics[width=\textwidth]{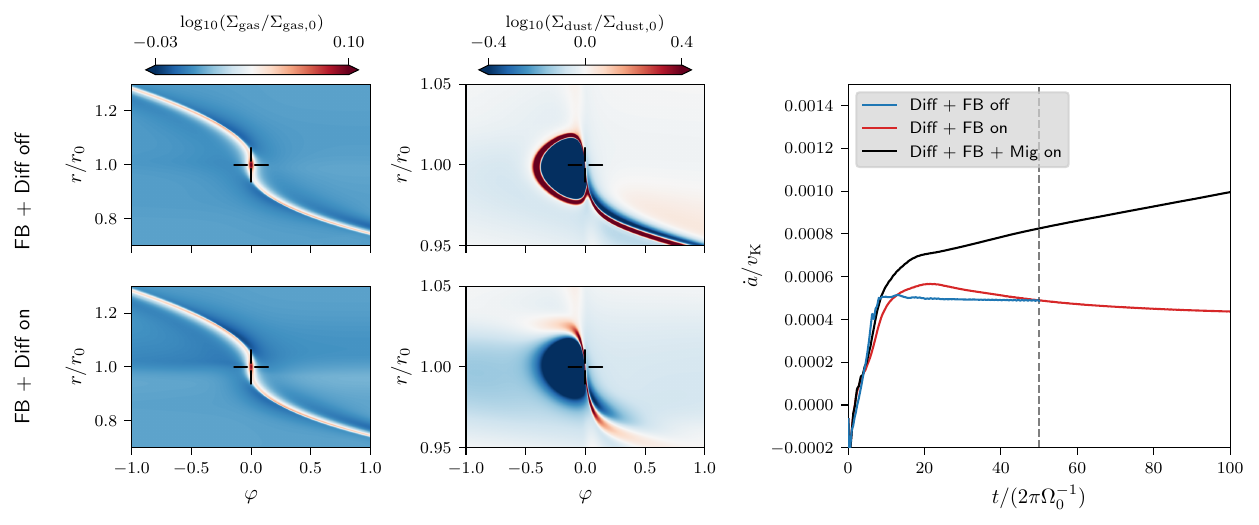} 
    \caption{Left and middle panels: relative surface density for the gas (left) and dust (middle) in the vicinity of a fixed planet after 50 orbits. Simulation parameters are $M_p = 4\,M_\oplus$, $\text{St} = 0.55$, $\epsilon = 0.05$, $\alpha = 10^{-3}$, and $\Sigma_0 = 5 \times 10^{-3}\,M_0/r_0^2$. The upper and lower rows isolate the cases where turbulent diffusion and dust feedback are turned off and on, respectively. Notably, the trailing dust void and the over-dense streaming filament ahead of the planet are preserved in both scenarios. 
    Right panel: time evolution of the dimensionless migration rate, $\dot{a}/v_\text{K}$. The blue and red lines correspond to the fixed-orbit scenarios displayed in the left and middle panels, respectively, while the black line represents a self-consistent simulation in which turbulent dust diffusion and feedback are enabled and the planet dynamically migrates due to net torque from the disk (gas + dust)  
    (S48 in Table \ref{tab:fullresult}). The vertical dashed line marks the 50-orbit epoch of the density snapshots. The fixed simulation with diffusion and feedback turned off reproduces results comparable to those of \cite{Benitez-Llambay2018}. While incorporating diffusion and feedback does not alter the overall direction of migration under fixed-orbit conditions, self-consistent migration noticeably modifies the migration rate, as evidenced by the departure between the red and black curves.}
    \label{fig:fig1}
\end{figure*}

Protoplanetary disks are not composed solely of gas. The solid component is the fundamental building block of planets.
While the role of dust in planetesimal formation and planetary growth is well-established, its direct influence on migration is yet to be understood. 
\cite{Benitez-Llambay2018} demonstrated through hydrodynamical simulations that the torque exerted by the dust component of a disk can be strongly positive and comparable in magnitude to the negative torque from the gas.

The key parameter governing this gas-dust coupling is the dimensionless Stokes number, ${\rm St}$. Physically, it represents the ratio of a particle's aerodynamic stopping time to the local orbital timescale, and measure how strongly a solid particle is coupled to the gas flow \citep[e.g.,][]{Chiang2010, Lambrechts2019}. For a given gas surface density, the Stokes number increases with particle size. However, rather than being determined solely by the background gas, the maximum ${\rm St}$ attainable by solids is limited by collisional fragmentation. As shown by \cite{Batygin2022}, this limit scales sublinearly with stellar distance and experiences a sharp two-order-of-magnitude jump across the water-ice sublimation front. Consequently, while silicate grains in the inner disk remain tightly bound to the gas flow (${\rm St} \sim 10^{-4}$), icy aggregates in the outer nebula reach ${\rm St} \sim 10^{-1}$, entering a marginally coupled regime.

To demonstrate the stark contrast in the response of these two species, Fig.\,\ref{fig:fig1} compares the steady-state density perturbations in the gas and dust components for a typical low-mass planet on a fixed orbit. 
The difference is striking. While the gas maintains a relatively smooth structure dominated by linear spiral waves that resemble an inner and outer spiral wake (left panels), the dust layer develops a strong morphological asymmetry (center panels). 
The planet clears a highly depleted cavity trailing its orbit, while a dense filament of solids resulting from the planet torquing the material that is close to it spirals inward.
This azimuthal asymmetry of solids is of paramount importance. The gravitational pull from this over-dense leading filament combined with the force imbalance resulting by the under-dense cavity, exerts a positive torque that can effectively counteract, and even overcome, the standard inward push from the gas.
This discovery has spurred further research into solid-driven migration, highlighting the need to consider the multiphase nature of disks \citep{Chen2018, Hsieh2020, Regaly2020, Hou2024, Hou2025, Chrenko2024, Chametla2025}. 
Furthermore, Fig.\,\ref{fig:fig1} illustrates the combined effect of dust feedback and turbulent diffusion on the solid distribution close to the planet (bottom row). We note that while turbulent diffusion causes the over-dense dust ring and filament to largely dissipate, the core asymmetry of the dust-void and leading filament is preserved. Interestingly, the static torques for both fixed-planet scenarios remain remarkably similar.

Understanding the mechanisms that govern this migration is crucial for explaining the observed distribution of exoplanets \citep[see e.g.,][]{Masset2001, Paardekooper2023}, and recent research suggests that dust-planet interactions can generate substantial positive torques, potentially altering the direction and speed of migration \citep{Guilera2023, Guilera2025}. However, a difference is observed when evaluating self-consistent migration (Fig.\,\ref{fig:fig1}, right panel, black line). While the fixed-orbit scenarios show no significant differences in the predicted migration rates, regardless of whether diffusion and feedback are included, allowing the planet to migrate as a consequence of the torque it feels results in a significant change. While prior models demonstrated the potential for dust to modify migration significantly, they lacked this self-consistent treatment of the planet's trajectory.

The key unanswered question remains: what happens when a planet is allowed to move and dynamically react to the very torques it generates? The planet's motion changes its position relative to the local dust distribution, which in turn alters the dust structure, thereby modifying the torque on the planet. This establishes a complex feedback loop that determines the ultimate fate of the planet, a mechanism that has not been fully characterized yet.

In this work, we close this loop by performing a comprehensive suite of self-consistent, two-fluid hydrodynamical simulations. By exploring a six-dimensional parameter space ($a_0$, $M_p$, ${\rm St}$, $\epsilon$, $\alpha$, $\Sigma_0$) through 144 distinct simulations, we characterize the global dynamics of low-mass cores. Crucially, we employ the comoving framework of \cite{Ramos2025} to reduce numerical diffusion of standard fixed-grid simulations of migrating planets, while speeding-up calculations. 
The main goal of this work is to systematically explore the diverse dynamical outcomes produced by the interplay between gas and dust torques during self-consistent migration. By mapping these trajectories, we seek to characterize the specific conditions under which low-mass planets undergo inward decay, outward drift, or become captured within self-induced stable equilibrium zones.

This paper is organized as follows. In Section \ref{sec:physical_model}, we outline our physical model and numerical setup. In Section \ref{sec:results}, we present the results of our parameter space exploration, demonstrating the existence of a variety of different migration regimes, including orbital confinement regions. In Section \ref{sec:discussion}, we present a discussion of our results, including caveats of our model and implications for planet formation. Finally, we summarize our findings and present our conclusions in Section \ref{sec:summary}.

\section{The physical model}
\label{sec:physical_model}
To study the problem of self-consistent migration in dusty disks, we perform two-dimensional multifluid hydrodynamic simulations. We solve the Navier-Stokes equations for a locally isothermal disk composed of gas and solids modeled as a pressureless fluid. The gas pressure is defined as $P = c_s^2 \Sigma_{\rm gas}$, where $c_s$ is the isothermal sound speed. While a locally isothermal equation of state simplifies the complex thermodynamics of the disks, this controlled approximation allows us to isolate the dynamical torques exerted by the dust component from thermal torques that arise when radiative effects are included. In this way we guarantee that the observed dynamics is exclusively driven by the dust-gas mechanical feedback.

We adopt a unit system such that our reference mass, distance, and time are $M_0=1$, $r_0=1$, and $\Omega_0^{-1}=1$, respectively. The disk is initialized following the NHS steady-state solution \citep{Nakagawa1986}, which accounts for the sub-Keplerian motion of the gas and the corresponding radial drift of the dust. The initial gas surface density profile is given by
\begin{equation}
    \Sigma_{\rm gas}(r) = \Sigma_0 \left(\frac{r}{r_0}\right)^{-1/2}\,,
\end{equation}
where $\Sigma_0 \in [6.37 \times 10^{-4}, 5.00 \times 10^{-3}] M_0/r_0^2$. We assume a constant aspect ratio $h \equiv c_{\rm s}/(r \Omega_{\rm K}) = 0.05$, with $\Omega_{\rm K} = \sqrt{GM_0/r^3}$ the Keplerian angular frequency.

We model turbulence using the standard $\alpha$-prescription \citep{Shakura1973}, exploring both low-viscosity, $\alpha=10^{-4}$, and high-viscosity, $\alpha=10^{-3}$, regimes. To isolate the effects of particle size, we consider a monodisperse dust-size distribution with a constant Stokes number varying ${\rm St} \in [0.10, 0.26, 0.55]$. The selected range of Stokes numbers corresponds to mm- to cm-sized icy pebbles in outer disk environments ($5\text{--}100\,\text{au}$), where ${\rm St} \sim 10^{-1}$ represents the marginally coupled regime introduced in Section~\ref{sec:introduction}. Lower Stokes numbers ($\mathrm{St} \le 0.1$) were excluded because tightly coupled dust passively follows the gas streamlines, suppressing the asymmetric co-orbital dust voids and filaments required to generate significant dust torques \citep{Benitez-Llambay2018}. In this regime, dust-driven torques become negligible and migration reverts to standard, gas-dominated Type-I behavior.
Although realistic protoplanetary disks are composed by solids whose size distribution is governed by ongoing coagulation and fragmentation, we adopt a monodisperse dust population with a fixed Stokes number to untangle the parameter space. The initial dust surface density is
\begin{equation}
    \Sigma_{\rm dust}(r) = \epsilon \Sigma_{\rm gas}(r)\,.
\end{equation}
We set $\epsilon = 0.01$ or $\epsilon = 0.05$, for low or high metallicity models, respectively. 
While the pressureless fluid approximation can artificially accumulate material due to its inability to capture orbit-crossing dynamics, the turbulent diffusion included in our models helps preventing nonphysical discontinuities at the grid scale.

Type-I migration theory shows that the torque exerted by a gaseous disk on a low-mass planet depends exclusively on local disk properties, such as the surface density and temperature gradient, and aspect ratio \citep[e.g.][]{Tanaka2002}. In this regime, the torque exerted by the disk, is
\begin{equation}
\Gamma_{\rm gas} = \psi \, \Gamma_{\rm p}\,,
\end{equation}
where $\psi$ is an $\mathcal{O}(1)$ factor that encapsulates the torque dependence with the disk surface density and temperature gradients. The reference torque, $\Gamma_{\rm p}$, is given by
\begin{align}
    \Gamma_{\rm p} =q^2h_{\rm p}^{-2}\Sigma_{\rm gas, p}a_{\rm p}^4\Omega_{\rm p}^2,
    \label{eq:torq_p}
\end{align}
where $ q = M_{\rm p} / M_\star$ is the planet-to-star mass and the subscript ${\rm p}$ denotes that the quantities must be evaluated at the planet's position. 
For a planet in circular orbit, its dimensionless migration rate due to the total torque exerted by the disk is \citep[e.g.,][]{Tanaka2002, Benitez-Llambay2026}:
\begin{align}
    \frac{1}{a_{\rm p}\Omega_{\rm p}} \frac{d a_{\rm p}}{dt} =  \frac{2 \Gamma_{\rm T} }{a_{\rm p}^2\,\Omega_{\rm p}^{2} }\,,
\end{align}
where $\Gamma_{\rm T} = \Gamma_{\rm gas} + \Gamma_{\rm dust}$. If we neglect the dust torque, we can find the scaling of the dimensionless migration rate with $a_{\rm p}$
\begin{align}
    \frac{1}{a_{\rm p}\Omega_{\rm p}} \frac{d a_{\rm p}}{dt} \simeq  \frac{2 \Gamma_{\rm p} }{a_{\rm p}^2\,\Omega_{\rm p}^{2} }\,.
\end{align}
After assuming a disk model in which the temperature is $T\propto (r/r_0)^{-1}$ ($h$ constant) and $\Sigma \propto (r/r_0)^{-\sigma}$, we obtain
\begin{align}
    \frac{1}{a_{\rm p}\Omega_{\rm p}} \frac{d a_{\rm p}}{dt} \propto \left(\frac{a_{\rm p}}{r_0}\right)^{-\sigma+2}\,.
\end{align}
Therefore, the dimensionless migration rate would remain constant if $\sigma=2$. However, for typical disk models and, in particular, a steady-state accretion disk ($\sigma = 1/2$), this condition is not met and the migration rate intrinsically depends on the planet's radial position. To map the orbital evolution and characterize the sensitivity of the trajectories on different disk locations, we systematically track planets migrating from inner ($ a_{\rm 0}/r_0 = 1$) and outer ($a_{\rm 0}/r_0 = 10$) disk locations.

\subsection{Numerical Method}
\label{sec:numerical_method}

We solve the two-fluid system using the publicly available hydrodynamical code \texttt{FARGO3D} \citep{Benitez-Llambay2016, Benitez-Llambay2019}. Additionally, our model incorporates turbulent diffusion \citep{Weber2019} and the self-consistent momentum exchange between gas and solids. We assume a dust diffusion model that scales directly with the turbulent viscosity of the gas. The dust diffusivity is parameterized as $D_{\rm d} = \delta c_s h r$, where $\delta$ represents the dimensionless diffusion parameter. For all simulations incorporating turbulent diffusion, we assume that solid diffusion is driven by the same turbulent eddies that provide the gas viscosity, setting $\delta = \alpha$. 
The planet mass takes $2.67$, $4$, and $10 M_{\oplus}$ and is allowed to migrate self-consistently from an initial circular and coplanar orbit around the central star of mass $M_\star = M_0$, responding dynamically to the combined torques from the gas and dust. We model the planet's gravity using a Plummer potential smoothed over one Hill radius ($R_{\rm H}$) and, as a conservative measure, exclude the torque contribution arising from the inner half of the Hill sphere ($0.5 R_{\rm H}$) to avoid artificial forcing from locally bound material \citep{Crida2009}. Given that we solve the equations in an astrocentric frame of reference, we include the contribution of the indirect potential in our simulations.

A critical feature of our model is the use of a comoving frame of reference  that keeps the migrating planet fixed within the mesh through the coordinate transformation $r' = r/a_{\rm p}$ \citep{Ramos2025}.
In standard fixed-grid simulations, a migrating planet moving radially through the cells is subject to enhanced numerical diffusion that smooths out sharp or subtle flow features. By adopting a framework that co-rotates and radially tracks the planet, we preserve the detailed dynamics in the planet's immediate vicinity. 
Furthermore, a comoving mesh eliminates tight timestep restrictions at the inner boundary, decrease the numerical cost per simulation. Because our simulations are performed in this comoving frame, we report our results using the comoving time $t'$, defined via the coordinate transformation \citep{Ramos2025}
\begin{equation}
    t' = \int \Omega_{\rm p} dt\,.
\end{equation}
In addition, in appendix \ref{apendix:equations}, we demonstrate that the diffusive flux remains strictly invariant under the comoving transformation. 

The computational domain spans radially from $r_{\rm min}/a_{\rm p} = 0.54$ to $r_{\rm max}/a_{\rm p} = 1.84$, discretized using a grid of $N_r = 1024$, $N_\varphi = 2048$ radial and azimuthal cells, respectively. The non-uniform grid follows the step functions defined by \cite{Benitez-Llambay2023} using $\varphi_a= 0.2$, $\varphi_b= 0.3$, $\varphi_c= 10$, $r_a= 0.05$, $r_b= 0.2$ and $r_c= 5$ \citep[$a$, $b$ and $c$ parameters of appendix B of][] {Benitez-Llambay2023}. This leads to a resolution of $H/dy \approx H/dx \approx 100$ in the uniform space region for a planet of $M_{\rm p} = 4 M_{\oplus}$ and a constant radial resolution per scale-height far from the planet as it migrates. 

To integrate the hydrodynamical equations on this non-uniform mesh without severe time-step penalties, we implement the Rapid Advection algorithm on arbitrary Meshes (RAM) \citep[][]{Benitez-Llambay2023}. RAM is a time-explicit, conservative operator-splitting scheme designed to solve advection in flows dominated by a large bulk velocity on arbitrary computational grids. By utilizing conservative piecewise monotonic reconstructions and a coordinate transformation based on a predefined mesh density function, RAM analytically shifts the advected profiles along the bulk flow direction. This effectively removes the bulk Keplerian velocity from the Courant-Friedrichs-Lewy (CFL) stability condition, allowing for significantly larger time steps and lower numerical diffusion compared to standard explicit upwind schemes. The combination of the comoving frame and RAM provides a significant numerical advantage. While the comoving frame ensures the planet semi-major axis constantly remains centered within the highest-resolution region of the mesh, RAM allows us to utilize a non-uniform grid without restrictive CFL limits. Together, they yield unparalleled precision in resolving the subtle, asymmetric co-orbital dust dynamics at a significantly lower computational cost.

Finally, to prevent artificial wave reflections at the radial boundaries, we implemented wave-killing zones near the inner and outer edges, relaxing the density and radial velocity fields towards their initial axisymmetric comoving values.


\section{Results}
\label{sec:results}

\begin{figure}[ht!]
    \centering
    \includegraphics[width=\linewidth]{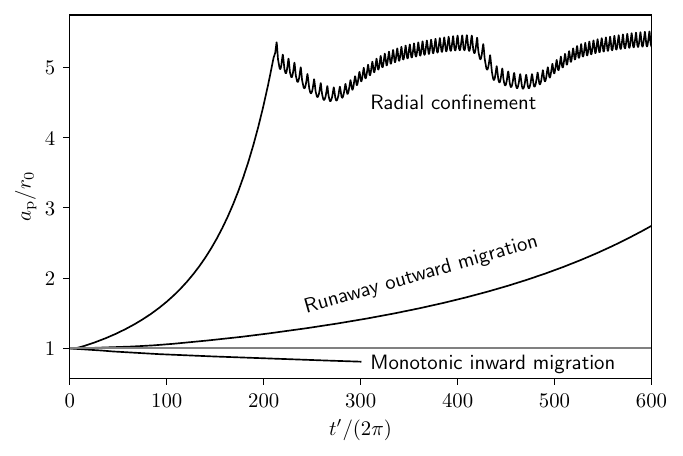}
    \caption{Evolution of the planet's semi-major axis, $a_{\rm p}$, as a function of comoving time, $t'/(2\pi)$, illustrating the three distinct deterministic migratory regimes. The bottom curve displays monotonic inward migration (S58 in Table\,\ref{apendix:table}), driven primarily by the gas torque. The middle curve shows monotonic outward migration (S18 in Table\,\ref{apendix:table}), resulting from a dominant positive dust torque. Finally, the top curve demonstrates oscillatory motion that results in radial confinement (S48 in Table\,\ref{apendix:table}).}
    \label{fig:fig3}
\end{figure}

\begin{table}[h!]
\centering
\begin{tabular}{cccccc}
\hline
$a_0/r_0$ & $M_p\,(M_{\oplus})$ & St & $\epsilon$ & $\alpha$ & $\Sigma_0\, (M_0/r_0^2)$ \\
\hline
1  & 2.67 & 0.10 & 0.01 & $10^{-3}$ & $5.00\times 10^{-3}$ \\
10 & 4.00 & 0.26 & 0.05 & $10^{-4}$ & $6.37\times 10^{-4}$ \\
 & 10.00 & 0.55 &  &  & \\
\hline
\end{tabular}
\caption{Summary of the six-dimensional parameters space explored in this work.}
\label{tab:6D-parameters}
\end{table}

\subsection{Migration regimes and parameter exploration}
\label{sec:global_regimes}

\begin{figure*}[th!]
    \centering
    \includegraphics[width=\textwidth]{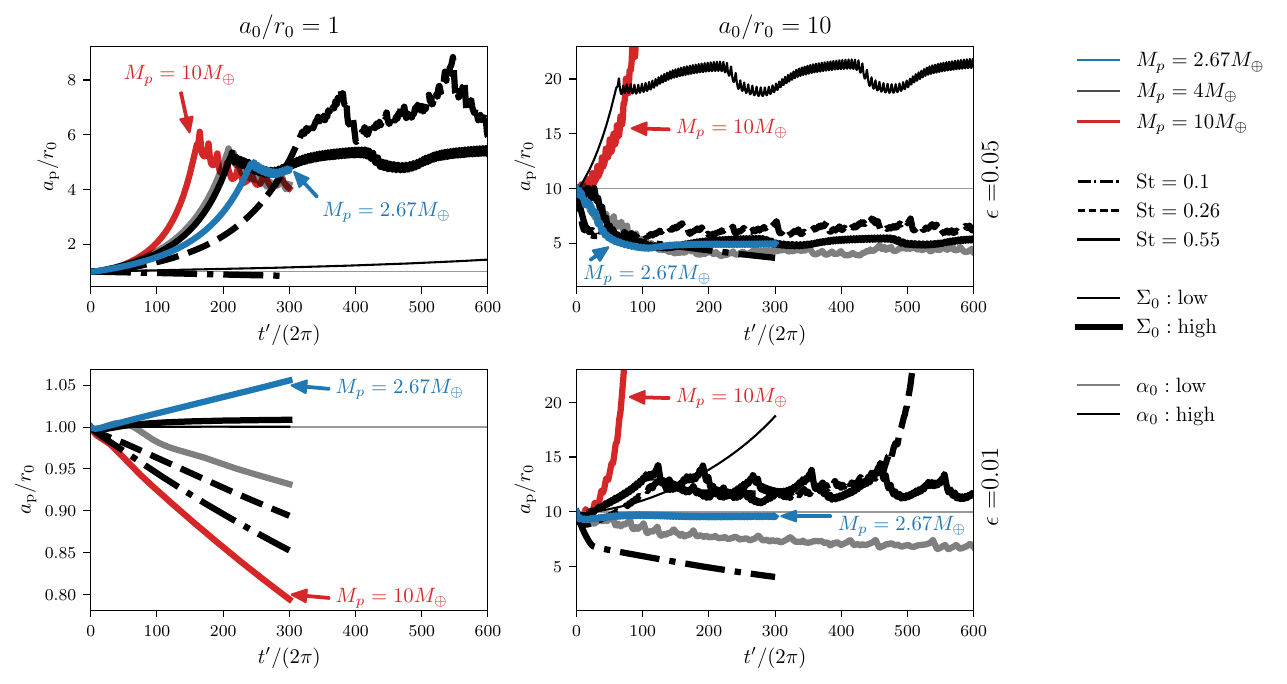}
    \caption{Time evolution of the planetary semi-major axis ($a/r_0$), showing a selected subset of migration regimes, mass bifurcation, and the emergence of multiple equilibrium zones. The panels isolate the effects of disk metallicity and initial orbital position ($a_0/r_0$). The top row corresponds to high-metallicity disks ($\epsilon=0.05$) and the bottom row to low-metallicity disks ($\epsilon=0.01$). Left and right columns separate initial positions $a_0/r_0=1$ and $a_0/r_0=10$, respectively. The reference simulation (S48: $M_p=4M_\oplus$, ${\rm St}=0.55$, $\epsilon=0.05$, $\alpha=10^{-3}$, high initial $\Sigma_0$) is indicated by the thick solid black line, demonstrating stable orbital trapping. Deviations from this baseline are encoded as follows: colors denote planetary mass (blue: $2.67M_\oplus$, black: $4M_\oplus$, red: $10M_\oplus$), line styles represent the Stokes number (solid: $\mathrm{St}=0.55$, dashed: $\mathrm{St}=0.26$, dashed-dot: ${\rm St}=0.1$), line thickness indicates initial gas surface density (thick: high $\Sigma_0$, thin: low $\Sigma_0$), and line transparency reflects turbulent viscosity (solid/dark lines: $\alpha=10^{-3}$, semi-transparent lines: $\alpha=10^{-4}$). The 28 simulations showed here are denoted with a dagger ($^\dagger$) in Table \ref{apendix:table}.}
    \label{fig:fig4}
\end{figure*}

To characterize the impact of the gravitational interaction between a planet and a dusty protoplanetary disk, we performed a comprehensive suite of 144 self-consistent simulations mapping orbital evolution across a six-dimensional parameter space ($a_0$, $M_p$, ${\rm St}$, $\epsilon$, $\alpha$, $\Sigma_0$) defined in previous section and summarize in Table\,\ref{tab:6D-parameters}. 
All our simulations are listed in Table\,\ref{apendix:table} of the appendix.

These simulations allowed us to identify three distinct migratory regimes, illustrated in Fig.\,\ref{fig:fig3}:

\paragraph{Monotonic inward migration} The planet follows a continuous orbital decay towards the central star. This occurs when the positive dust torque is insufficient to overcome the negative torque from the gas, leading to standard or reduced Type-I migration.
In some cases, the dust torque is also negative, further accelerating the inward migration driven by the gas component.

\paragraph{Monotonic/runaway outward migration} The planet exhibits sustained outward migration, driven by a highly dominant positive dust torque. This outward migration is found to be runaway in some cases.

\paragraph{Radial confinement} The planet is kept within a radial band where the total torque vanishes on average. Within this zone, the planet semi-major axis exhibits bounded, stable oscillations, effectively halting orbital decay.

\vspace{1em}

In Fig.\,\ref{fig:fig4} we present these migration regimes. We show the time evolution of the planet's semi-major axis for a representative subset of our parameter grid. 
The four panels are designed to isolate the effects of disk metallicity and the initial semi-major axis, with specific physical parameters denoted by distinct line colors and styles. We define our reference simulation (indicated by the thick solid black line, simulation S48 in Table\,\ref{apendix:table}) using the following parameters: $M_p=4M_\oplus$, ${\rm St}=0.55$, $\epsilon=0.05$, $\alpha=10^{-3}$, and a high initial surface density ($\Sigma_0$). By analyzing these trajectories, we can identify how each parameter dictates the planet's migratory fate:

\paragraph{Planet mass} The planetary mass fundamentally controls two dynamical features: the migration velocity (driven by the balance between gas and dust torques) and the amplitude and/or stability of the radial oscillations. As seen in Fig.\,\ref{fig:fig4}, the mass governs the steepness of the initial migration tracks. In high-metallicity environments ($\epsilon=0.05$), the dust positive torque dominates, propelling massive cores ($10M_\oplus$) outward at significantly higher velocities than their lower-mass counterparts ($2.67M_\oplus$ and $4M_\oplus$). Conversely, in low-metallicity disks ($\epsilon=0.01$, bottom-left panel), the dust torque is not enough to counteract the negative gas torque, forcing the $10M_\oplus$ core to migrate rapidly inward while the lower-mass planets manage to maintain a slow outward drift. Furthermore, planetary mass regulates the stability of the confinement zones. Lower-mass planets ($2.67M_\oplus$) transition into highly stable traps characterized by tight radial oscillations. As the mass increases to $4M_\oplus$, the amplitude of these oscillations grows noticeably. For massive cores ($10M_\oplus$), the solid-driven interactions become strong, inducing orbital eccentricity until the trap is destabilized, preventing stable long-term confinement and leading to either runaway trajectories or ejection from the simulation domain.

\paragraph{Metallicity} The disk metallicity dictates the global strength of the solid-driven torques. It also determines whether orbital traps act as deterministic attractors or transient states. In high-metallicity disks ($\epsilon=0.05$, top panels), positive dust torques are larger than gas torques, driving super-Earths into stable oscillatory confinement and triggering runaway outward migration for massive cores. Conversely, in low-metallicity environments ($\epsilon=0.01$, bottom panels), the orbital evolution becomes highly sensitive to parameters. Rather than a universal inward decay for all planets, Fig.\,\ref{fig:fig4} reveals a complex mass-dependent inversion. At $a_0=1$ (bottom-left), a massive $10M_\oplus$ core undergoes rapid monotonic inward migration because the depleted dust asymmetry completely fails to counteract the strong negative gas torque. In stark contrast, the lowest-mass embryos ($2.67M_\oplus$ and $4M_\oplus$) can maintain a slow outward drift or hold their orbital position, provided $St=0.55$ in a highly viscous disk ($\alpha=10^{-3}$). While this is in agreement with recent findings by \cite{Chametla2025}, who reported that dust fails to generate a net positive torque under low turbulence ($\alpha=10^{-4}$), our expanded grid reveals that high viscosity paired with large particles can induce solid-driven outward migration even at low metallicity. Finally, for planets starting in the outer disk ($a_0=10$, bottom-right), low metallicity renders the confinement zones more fragile and less defined, typically replacing the robust limit cycles seen at $\epsilon=0.05$ with either less deterministic trapping, unconfined scattering or monotonic inward/outward drift.

\paragraph{Aerodynamic Coupling} The aerodynamic coupling between the gas and solid components affects the efficiency of the dust torque. The line styles in Fig.\,\ref{fig:fig4} explicitly reveal how this coupling dictates planet migration. Small particles (${\rm St} = 0.1$, dashed-dot lines) result in weaker dust asymmetries, failing to counteract the negative gas torque, resulting in inward migration across nearly all tested disk environments, with just a few cases (7 out of 48) of trapping and outward migration (see S6, S77, S78, S101, S102, S126, and S128 of Table\,\ref{apendix:table} in the appendix). In contrast, larger, marginally coupled particles (${\rm St} = 0.55$, solid lines) easily decouple from the gas and create asymmetries that exert significant positive torque. These particles generate the strongest positive torques, driving rapid outward migration and locking planets into the most robust oscillatory traps of our exploration (e.g., the rapid convergence of the solid blue and black lines to $a_{\rm p}/r_0 \approx 5$ in the top-left panel). Even in low-metallicity environments (bottom-left panel), ${\rm St} = 0.55$ dust is capable of driving outward migration for super-Earths. The intermediate coupling regime (${\rm St} = 0.26$, dashed lines) represents a transitional state. While these particles can drive outward migration and trapping in high-metallicity disks, their trajectories exhibit slower initial outward velocities and significantly wider, more erratic radial oscillations. Furthermore, for low-metallicity disks, the positive torque produced by these particles is typically smaller than the negative gas torque, resulting in inward migration for $a_0/r_0=1$. In some cases, the dust torque is also negative, further accelerating the inward migration driven by the gas component. For low-mass planets ($2.67M_\oplus$), this negative torque is weak; because these small grains are strongly coupled aerodynamically, the weak planetary gravity cannot break this coupling to clear a co-orbital cavity, leaving the dust to trace the gaseous spiral wakes (see S6, S1, S2, S3 and S4 of Table\,\ref{apendix:table} in the appendix). In contrast, for high-mass planets ($10M_\oplus$), the negative dust torque becomes highly significant. Instead of preserving a depleted co--orbital cavity behind the planet—which under other parameters remains mostly empty and allows the positive torque that halts migration—-the strong planetary gravity distorts co-orbital flow. This makes the dust torque negative, cooperating with the gaseous component to accelerate inward orbital decay toward the host star (see S49, S50, S51, and S52 of Table\,\ref{apendix:table} in the appendix). Finally, for high-metallicity, these particles can drive outward migration and trapping.

\paragraph{Surface density} The background gas surface density ($\Sigma_0$), indicated by line thickness in Fig.\,\ref{fig:fig4}, acts as a regulator for both the spatial location of the attractors and their capture efficiency. Massive disks (thick lines) favour the equilibrium zones in the inner disk, typically capturing planets between 4 and 7 $r_0$. In stark contrast, less dense disks (thin lines) systematically shift the net-zero torque outward, establishing wide-orbit traps often located between 14 and 27 $r_0$ (see S95, S111, S113, and S119 in Table\,\ref{apendix:table}).  For embryos migrating inward from the outer disk ($a_0/r_0 = 10$), high $\Sigma_0$ is strictly required to successfully halt migration. Nearly every inward-to-trapped transition in our grid corresponds to a high-density environment (e.g., S96, S112, and S120). Conversely, for low-mass planets that migrate outward from the inner disk ($a_0/r_0 = 1$), low surface densities generally fail to bound the orbit, leading to outward migration rather than settling into a stable limit cycle, a bifurcation explicitly captured when comparing the continuously outward simulation S47 (low $\Sigma_0$) to the tightly trapped simulation S48 (high $\Sigma_0$).

\paragraph{Turbulent viscosity} The level of turbulent viscosity ($\alpha$), indicated by line opacity in Fig.\,\ref{fig:fig4}, dictates the stability of the attractors and regulates the exact radial location of the traps. Higher viscosity (opaque lines, $\alpha=10^{-3}$) creates well-defined radial oscillations with large, regular amplitudes. In this high-viscosity regime, the strong diffusion coupled with dust feedback allows the leading filament to continually regenerate. Conversely, lower diffusivity (transparent lines, $\alpha=10^{-4}$) produces more irregular, high-frequency, chaotic oscillations without clear periodicity. Furthermore, lower viscosity systematically shifts the time-averaged position of the traps inward. A quantitative comparison from our set of simulations (Table\,\ref{apendix:table}) confirms this spatial shift. An identical $4 M_\oplus$ embryo migrating from the outer disk ($a_0/r_0=10$) in a dense, high-metallicity environment is trapped at $a \approx 5.16$ $r_0$ under high viscosity (S120), but is pushed deeper into the disk, halting at $a \approx 4.31$ $r_0$, under a low viscosity configuration (S118). A parallel inward shift is observed for outward-migrating embryos from the inner disk (comparing S48 and S46, where the trap shifts inward from 5.16 $r_0$ to 3.35 $r_0$). Finally, high viscosity can enhances planetary survival in marginal environments. For example, in low-metallicity disks (bottom-left panel), increasing $\alpha$ to $10^{-3}$ allows to amplify the dust torque sufficiently to reverse the inward migration seen at $\alpha=10^{-4}$ (e.g., outward drift of S44 against the rapid inward decay of S42 represented by the black solid curves).

\paragraph{Initial position} The planet initial semi-major axis ($a$) acts as a bifurcation parameter in our exploration, revealing a transition region where global disk properties alone are insufficient to predict the orbital evolution of the planet. Comparing the left ($a_0/r_0=1$) and right ($a_0/r_0=10$) columns of Fig.\,\ref{fig:fig4} shows different dynamical responses. For embryos originating in the inner disk, trapping is highly restrictive: among the 72 simulations initialized at $a_0/r_0=1$, the majority (46 cases) result in monotonic inward migration. While 26 simulations exhibit outward migration, only 11 successfully settle into a stable oscillatory regime. Crucially, this inner-disk confinement requires both high gas surface density ($\Sigma_0 = 5 \times 10^{-3}$) and high metallicity ($\epsilon = 0.05$). Conversely, migration from the outer disk ($a_0/r_0=10$) presents a more complex interplay. Because the background gas density is intrinsically lower at wide orbits ($\Sigma \propto r^{-1/2}$), the balance between gas and dust torques shifts, enabling trap survival under a broader range of conditions. For instance, an identical $4 M_\oplus$ embryo with weak aerodynamic coupling (${\rm St}=0.1$) in a low-turbulence ($\alpha=10^{-4}$), high-density metal-rich disk undergoes monotonic inward migration when starting at $a_0/r_0=1$ (see S30 in Table\,\ref{apendix:table}), but is successfully halted and captured by a dynamical attractor near 5 $r_0$ when starting at $a_0/r_0=10$ (S102). Of the 41 outer-disk simulations that initially migrate inward, 12 are successfully captured by traps in dense, metal-rich environments. Furthermore, the outer disk allows for reverse capture, where 13 cases are trapped while initially migrating outward. Unlike the inner disk, all but one of these captures occur in low surface density environments, indicating that a lower gas density facilitates the reversal of the outward migration direction before trapping occurs. However, this reduced gas damping in the outer disk also permits severe dynamical instabilities. We identify a subset of 11 special cases that experience extremely rapid outward migration, encounter transient stability regions where they oscillate briefly, and subsequently undergo rapid inward reversals or rapid eccentricity growth (see the asterisk-marked simulations in Table\,\ref{apendix:table}). A detailed analysis of these unstable trajectories is beyond the scope of the present study and will be the subject of future work.

\subsection{Dynamical attractors as migration traps}
\label{sec:mechanics}

Having established the existence and distribution of orbital traps across the parameter space, we now turn to the underlying physical mechanisms that govern their dynamics. 

\begin{figure}[t!]
    \centering
    \includegraphics[width=\linewidth]{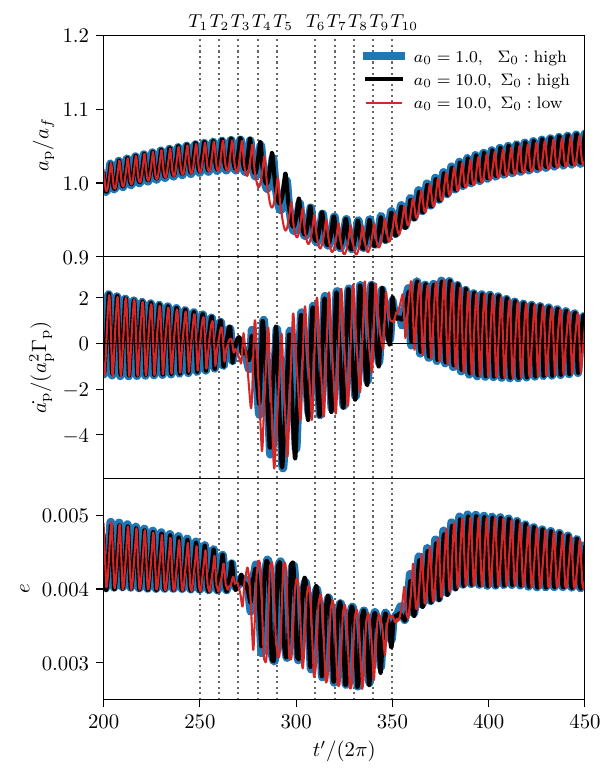} 
    \caption{Time evolution of the planetary semi-major axis normalized by its final time-averaged equilibrium value ($a_{\rm p}/a_f$) (top), dimensionless migration rate (middle), and eccentricity (bottom) for a $4 M_{\oplus}$ planet in a highly viscous, metal-rich disk ($\alpha = 10^{-3}$, $\epsilon = 5.0 \times 10^{-2}$, ${\rm St} = 0.55$). A uniform filter was applied to the data in all three panels every 10 orbits. To demonstrate the universal scalability of the dynamical attractor, the curves have been phase-shifted. This shift and normalization explicitly reveals that, despite concurring at different absolute orbital distances and integration times, the dust-modulated traps converge into an identical oscillatory limit cycle. The vertical gray dotted lines, labeled $T_1$ through $T_{10}$, mark the exact comoving times ($t'$) cross-sectioned in Figs.\,\ref{fig:fig6} and \ref{fig:fig7}. These three simulations are indicated with gray rows in Table\,\ref{apendix:table}.}
    \label{fig:fig5}
\end{figure}

\subsubsection{Radial trapping and universal scaling}
A fundamental result of this work is the phenomenon of radial convergence. As shown in our previous analysis, low-mass planets ($M_{p} \le 4 M_{\oplus}$) migrating outwards from the inner disk ($a_0/r_0 = 1$) and inwards from the outer disk ($a_0/r_0 = 10$) often converge to the exact same radial location, for example near $a \approx 5$ $r_0$. 

\begin{figure}[t!]
    \centering
    \includegraphics[width=\linewidth]{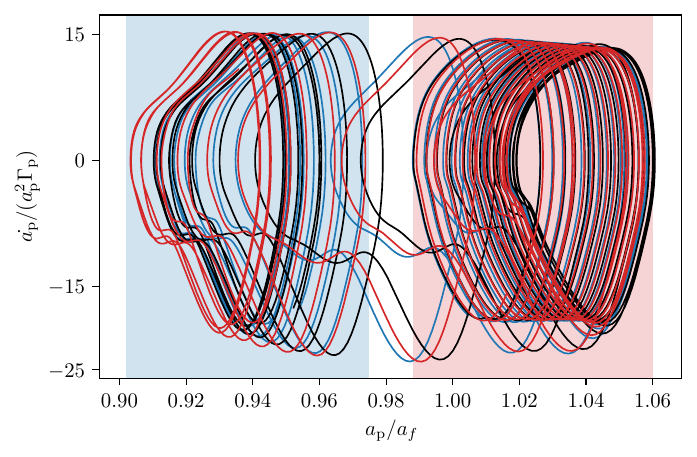}
    \caption{Phase-space of the dynamical attractors corresponding to the simulations in Fig.\,\ref{fig:fig5}. By mapping the dimensionless migration rate against the normalized semi-major axis, the emergence of distinct oscillatory limit cycles is clear. The blue and red regions highlight the stable confinement phases for the periods between 200--288 and 295--300 orbits, respectively. The sparse trajectories connecting these regions trace the dynamical transition as the planet shifts between equilibrium centers.}
    \label{fig:phase-space}
\end{figure}

To unveil the underlying universal behavior of this confinement, Fig.\,\ref{fig:fig5} displays the temporal evolution of the semi-major axis, the migration rate and the eccentricity, normalized by suitable quantities. In particular, the semi-major axis has been normalized by its final time-averaged equilibrium value ($a_{\rm p}/a_f$). By filtering out the absolute spatial scale of each independent simulation, this normalization collapses the trajectories into an identical, self-similar oscillatory limit cycle. The phase shift between the semi-major axis and the migration velocity suggests that the trap operates as a phase-space attractor (see Fig.\,\ref{fig:phase-space}). The acceleration and deceleration phases map onto one another regardless of the planet's initial location. This provides evidence that the trap operates under a universal physical scaling dictated by local hydrodynamic conditions, and that it is independent of the previous orbital history. 

In addition to the convergence zone near 5 $r_0$, simulations with a high dust-to-gas mass ratio ($\epsilon=0.05$) and an initial orbital radius of $a_0/r_0=10$ reveal a second confinement zone located near $a\sim 20$ $r_0$, where the semi-major axis evolution flattens after an initial outward drift. The emergence of these multiple equilibrium regions is a consequence of the disk radial gradients. Because the background gas surface density ($\Sigma \propto r^{-1/2}$) and temperature intrinsically decay with orbital distance, the local aerodynamic coupling and the baseline gas torques vary continuously. This radial stratification creates multiple, distinct orbital radii where the positive dust torque perfectly cancels out the local negative gas torque, resulting in multiple zero-torque crossings that act as isolated dynamical attractors.

It is crucial to note that the long-term viability of these dynamical attractors is also tied to the planet eccentricity. As shown in the bottom panel of Fig.\,\ref{fig:fig5}, the eccentricity oscillates together with the semi-major axis, generally remaining bounded during stable oscillatory confinement. The eccentricity systematically peaks during the outward migration phase, driven by the strong, asymmetric gravitational torque from the over-dense dust filaments, and subsequently damps as the migration direction reverses. However, the continuous forcing from this dust asymmetry acts as a periodic pump. In certain parameter regimes—particularly for massive planets or in environments with marginal aerodynamic coupling—this solid-driven interaction induces a gradual growth in eccentricity. As recorded in our set of simulations (see the asterisk-marked simulations in Table\,\ref{apendix:table}), once the eccentricity surpasses a critical threshold ($e \gtrsim 0.2$), the planet's radial excursions become large, destabilizing the co-orbital dust structure and breaking the confinement. It is worth noting that due to the limited extent of the comoving simulation box, once the eccentricity becomes large, the spiral wakes launched by the planet do not fit within the domain, compromising the precision of the gravitational force exerted onto the planet. Studying these highly eccentric cases is beyond the scope of our work, but they should be investigated further in future studies by means of tailored setups (for example, by replacing $a_{\rm p}$ with the instantaneous orbital radius $r_{\rm p}$ in the comoving transformation and applying suitable boundary conditions).

A closer inspection of Fig.\,\ref{fig:phase-space} reveals an important physical property of the oscillatory traps. The trajectories do not form a perfectly thin, idealized limit cycle typical of simple systems. Instead, the attractors exhibit a characteristic thickness or phase-space diffusion. This variance could be a reflect of complex non-linear fluid flow. Small fluctuations within the gas and orbital variations in the exact morphological reconstruction of the leading dust filament continuously perturb the instantaneous torque. However, the fact that these continuously perturbed paths remain strictly bounded within specific topological regions—despite the stochastic noise of the fluid—demonstrates the robustness of the underlying dynamical attractor. 
The robust, quasi-periodic nature of these trajectories suggests that the complex hydrodynamics governing the orbital traps could be abstracted into a reduced analytical framework. Because the confinement fundamentally relies on the temporal mismatch between the planet's instantaneous velocity and the restructuring timescale of the co-orbital dust, this system could be modeled via a non-linear driven oscillator. In such a toy model, the local dust torque could be parameterized with an explicit time-delay component to simulate the physical lag in the hydrodynamic replenishment of the leading filament. Developing such a 1D analytical model would allow for rapid, computationally inexpensive explorations of the parameter space and will be a highly valuable avenue for future theoretical studies.

\subsubsection{Restoration of dust asymmetry}

\begin{figure*}[ht!]
    \centering
    \includegraphics[]{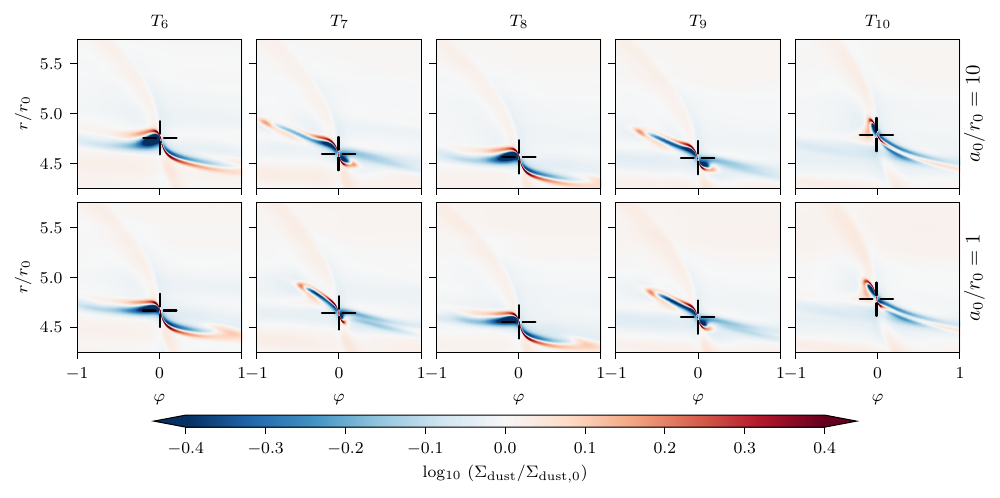} 
    \caption{Temporal sequence of relative dust surface density maps ($\log_{10}(\Sigma_{\text{dust}}/\Sigma_{\text{dust},0})$), displayed across five representative phases of the limit cycle ($T_6$ through $T_{10}$), corresponding to the last five vertical dotted lines in Fig.~\ref{fig:fig5}, illustrating the hydrodynamic restoration of the dust trap during an inward-to-outward migration reversal for a $4\,M_\oplus$ planet. Both rows share identical disk parameters ($\alpha=10^{-3}$, $\epsilon=0.05$, $St=0.55$, and high $\Sigma_{0}$), differing only in their initial position. Top Row: A planet initialized at $a_0/r_0=10$ being captured after its initial inward migration. Bottom Row: A planet initialized at $a_0/r_0=1$ undergoing a similar bounce within the limit cycle. As the planet's inward velocity slows, local conditions allow the solid material to replenish, dynamically restoring the sharp leading dust filament and trailing cavity. This cyclic restoration reactivates the strong positive torque necessary to halt inward decay.}
    \label{fig:fig6}
\end{figure*}

\begin{figure*}[ht!]
    \centering
    \includegraphics[]{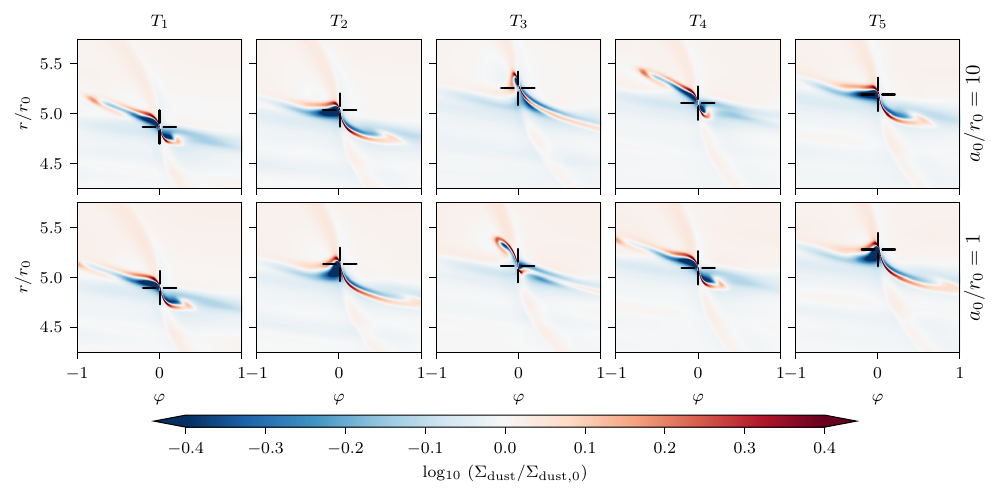} 
    \caption{Temporal sequence of relative dust surface density maps ($\log_{10}(\Sigma_{\text{dust}}/\Sigma_{\text{dust},0})$), 
    displayed across five representative phases of the limit cycle ($T_1$ through $T_5$), corresponding to the first five vertical dotted lines in Fig.~\ref{fig:fig5}, illustrating the dust structure during an outward-to-inward migration reversal within the trap for a $4\,M_\oplus$ planet. As in Fig.~\ref{fig:fig6}, both panels use the exact same disk setup, varying only the initial position. Top Row: A planet initialized at $a_0/r_0=10$ undergoing an outward-to-inward bounce. Bottom Row: A planet initialized at $a_0/r_0=1$ reaching the outer limit of its initial outward expansion. 
    During rapid outward migration, the planet's orbital velocity outpaces the local hydrodynamic replenishment of the solid material. The leading dense filament gets weaker, causing the positive dust torque to be insufficient to counteract the negative gas torque, forcing the migration direction to reverse inward.}
    \label{fig:fig7}
\end{figure*}

To understand the physical mechanism driving this radial confinement, we analyze the structural evolution of dust close to the migrating planet. Rather than being produced by a static disk feature, this oscillatory limit cycle is governed by the dynamical interplay between the planet's orbital velocity and the restructuring mechanism of the surrounding dust. The planet and its self-induced dust trap are effectively inter-locked in a continuous dynamical process. 
This delicate interplay is explicitly captured in the temporal sequences of the dust surface density. Consider first the outward-to-inward migration reversal illustrated in Fig.\,\ref{fig:fig7} for a $4\,M_\oplus$ planet (phases $T_1$ through $T_5$). During the outward migration phase, a characteristic morphological asymmetry emerges. This corresponds to a highly depleted cavity trailing the planet, while a dense filament of solids forms in the leading co-orbital region, generating the positive torque necessary to drive the outward drift. However, as the planet accelerates its outward migration, its radial velocity eventually outpaces the local replenishment rate of solid material. Because the dust dynamics is governed by its own aerodynamic coupling timescale, the solids cannot continuously restructure fast enough to maintain the asymmetry. As seen progressing through the rightmost panels of Fig.\,\ref{fig:fig7} (phases $T_4$ through $T_5$), the leading dense filament becomes diffuse. As this critical asymmetry washes out, the positive dust torque drops below the threshold of the permanent negative gas torque, forcing the migration direction to reverse inward.

Conversely, Fig.\,\ref{fig:fig6} illustrates the restoration of the trap during the subsequent inward-to-outward bounce for these same $4\,M_\oplus$ cores (phases $T_6$ through $T_{10}$). During the initial stages of inward migration, the dust structure momentarily tends toward axisymmetry, which minimizes the dust torque and allows the negative gas torque to dominate. Yet, as the planet's inward velocity gradually slows down, the local hydrodynamic timescale once again align with the planetary motion. This deceleration is caused from the rapid radial drift, which dynamically perturbs the gas and weakens the negative gas torque, providing the temporal window necessary for the solid material to accumulate again. As observed across the snapshots in Fig.\,\ref{fig:fig6}, the kinematic conditions permit the leading dust asymmetry to dynamically rebuild. This structural restoration reactivates the strong positive torque, effectively stopping inward decay and propelling the embryo back outward, restarting the limit cycle indefinitely.

\section{Discussion}
\label{sec:discussion}

\subsection{Self-Induced trapping}

Overcoming the rapid inward drift of planetary embryos has historically relied on specific structural or thermodynamic features within the background disk. Traditional solutions typically invoke either gas pressure maxima, such as dead-zone boundaries or gap edges \citep[e.g.,][]{Masset_2006, Morbidelli2020, Guilera2020, Lau2022, Sandor2024}, or specific thermal effects that generate strong positive corotation torques and create zero-torque islands \citep[e.g.,][]{Paardekooper2011, Benitez-Llambay2015, Masset2017}. While these thermal traps are effective, their existence and radial locations are highly sensitive to the disk's opacity, local cooling timescales, and background thermodynamic gradients. In this work, by presenting the first study of self-consistent migration of low-mass planets in fully dust-gas coupled disks, we demonstrate the viability of a fundamentally distinct mechanism, a self-induced orbital trapping driven entirely by the dynamics and distribution of the solid component around the planet.

Our results diverge fundamentally from the static fixed-planet approach initially presented by \citet{Benitez-Llambay2018} and subsequently applied in evolutionary models by \citet{Guilera2023, Guilera2025}. While fixed-orbit models can successfully identify regions of positive dust torque, allowing for self-consistent migration captures the complex dynamics of the co-orbital mass and the continuous dust-gas momentum exchange. This renders the resulting torques intrinsically dependent on the planet instantaneous velocity. We establish that robust orbital confinement can be sustained entirely by the local dust-planet interaction, without the need to invoke pre-existing pressure bumps or rely on strict thermodynamic gas conditions. The azimuthal asymmetry and the leading dust filament generated by the low-mass planet itself can act as a dynamical attractor where, in average, the planet does not migrate. Ultimately, even if a planet undergoes rapid initial inward migration from the outer disk, this self-induced attractor can effectively capture the embryo and bound its trajectory, offering a deterministic solution to planetary survival.

\subsection{Dust feedback vs. turbulent diffusion}

\begin{figure*}[th!]
    \centering
\includegraphics[width=\textwidth]{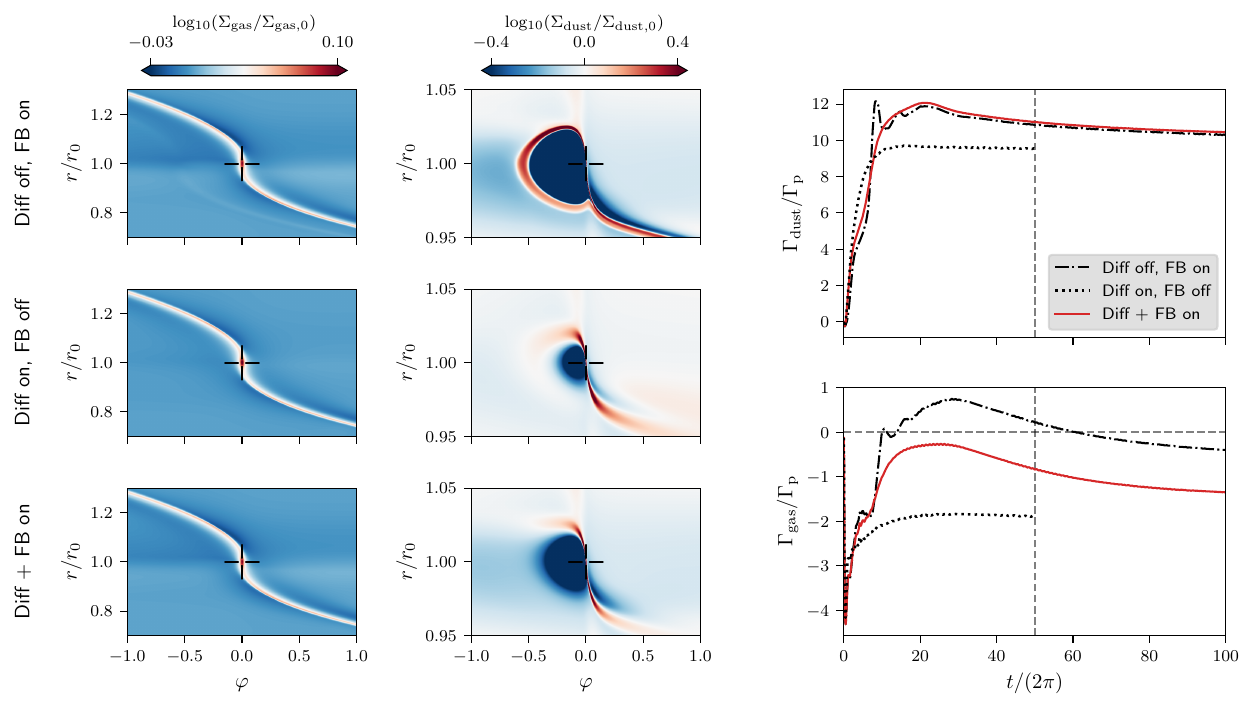}
\caption{Left and middle panels: Relative surface density perturbations for the gas (left) and dust (middle) after 50 orbits. Simulation parameters are identical to those in Fig.\,\ref{fig:fig1}. The rows isolate distinct physical mechanisms: dust feedback on, turbulent diffusion off (top), where the dust asymmetry is sharpest; dust feedback off and turbulent diffusion on with $\delta = \alpha$ (middle), where diffusion smooths the dust void and reduces the peak density of the leading filament, and the full model incorporating both feedback and turbulent diffusion (bottom). Comparing these panels reveals that dust feedback helps counteracting the effect of diffusion, maintaining a positive dust torque comparable to the lower viscosity case while keeping the negative gas torque lower in magnitude. Right panels: Time evolution of the torques normalized by $\Gamma_{\rm p}$ (see Eq.\,\ref{eq:torq_p}) for the dust (top) and gas (bottom). The inclusion of feedback (dash-dotted and red lines) consistently yields a higher positive dust torque and a less negative gas torque (closer to zero) compared to the case without feedback (dotted line), overall favoring outward migration.}
\label{fig:fig2}
\end{figure*}

Previous studies exploring planetary migration in dust-rich disks have raised valid concerns regarding the robustness of solid-driven torques. For instance, \citet{Hsieh2020} reported that dust-driven vortices can induce highly stochastic migration, rendering long-term orbital evolution chaotic and unpredictable. Similarly, recent work by \citet{Chametla2025} demonstrated that turbulent diffusion effectively smooths co-orbital dust structures, leading them to conclude that dust asymmetries are typically suppressed when diffusivity exceeds a critical threshold ($\delta \gtrsim 3 \times 10^{-4}$). In such regimes, solid-driven effects are predicted to be degraded or absent.
To explain how our self-induced traps survive as dynamical attractors, we must isolate the microscopic competition between the stabilizing force of dust feedback and the destructive nature of turbulent diffusion. By utilizing a fiducial fixed-orbit configuration ($M_{p} = 4 M_{\oplus}$, ${\rm St} = 0.55$, $\epsilon = 0.05$, $\alpha = 10^{-3}$, $\Sigma_0= 5\times 10^{-3}\,M_0/r_0^2$) and systematically decoupling the torque contributions (see Figs.\,\ref{fig:fig1} and \ref{fig:fig2}), we validate these concerns under specific limits. When turbulent diffusion is introduced without the back-reaction of dust onto the gas (Fig.\,\ref{fig:fig2}, middle row), it rapidly smooths sharp density gradients, washing out the trailing cavity and reducing the peak surface density of the leading filament. Consequently, the positive dust torque is reduced (Fig.\,\ref{fig:fig2}, right panel, dotted line), in agreement with \citet{Chametla2025}.
However, our full model reveals a different outcome driven by the active role of dust-gas feedback. When self-consistent feedback is activated alongside turbulent diffusion (Fig.\,\ref{fig:fig2}, bottom row), the back-reaction of the dense solid component onto the gas actively counteracts the diffusive smoothing. Even operating at a high viscosity of $\alpha = 10^{-3}$—well above previously established critical diffusivity thresholds—a high local metallicity ($\epsilon = 0.05$) helps maintaining a total positive torque. This effect not only allows the leading dust filament to persist and dynamically regenerate, but the local back-reaction also alters the surrounding gas distribution. As explicitly shown in the torque evolution (Fig.\,\ref{fig:fig2}, right panels, red and dash-dotted lines), this feedback mechanism consistently amplifies the positive dust torque while simultaneously rendering the negative gas torque closer to zero.
Consequently, rather than resulting in a chaotic random walk predicted by previous models, the back-reaction force shields the co-orbital asymmetry from diffusion. This guarantees that the system reliably evolves toward a bounded, deterministic oscillatory limit cycle, establishing dust feedback as a powerful mechanism for securing planetary survival even in highly diffusive disk environments.

\subsection{Mass bifurcation and dimensionality caveats}

Our parameter space exploration reveals a strong dependence on planetary mass, identifying a bifurcation near $M_{p} \approx 10 M_{\oplus}$. While super-Earths ($M_{p} \le 4 M_{\oplus}$) successfully transition into stable oscillatory confinement, massive cores in dust-rich environments eventually overcome the dust trap and enter a runaway outward migration regime. 
In these cases, the system enters a dynamical state analogous to gas-driven Type III migration \citep{masset2003}, but the threshold for this runaway regime is heavily mediated by the solid component.
Notably, while recent 3D simulations by \citet{Chametla2025} reported runaway inward migration for planets as small as $1.5 M_{\oplus}$, our self-consistent 2D framework indicates that this regime becomes dominant primarily for much more massive planets. As highlighted by \citet{Hsieh2020}, this quantitative difference shows the impact of dimensionality on the competition between turbulent diffusion and dust feedback in the context of different disk geometries. In a purely 2D framework, the density is concentrated within a single midplane, which maximizes the local dust density and amplifies the subsequent gas-dust feedback onto the planet.
Consequently, the resulting leading filament and trailing cavity in our 2D models are more pronounced. This amplified local dust concentration provides a massive positive torque that can resist the classical gas-driven inward push. Because this force is overestimated in 2D, we expect that a significantly larger planetary mass would be required to overcome the trap by breaking the solid asymmetry. In contrast, fully stratified 3D models account for vertical dust settling and turbulent diffusion, which would naturally diffuse the density at the midplane. This vertical diffusion results in a comparatively weaker leading asymmetry, explaining why 3D models observe runaway inward migration at lower planetary masses.
In agreement with finding by \cite{Chametla2025}, we therefore expect that the precise mass boundaries separating oscillatory confinement from runaway inward migration would shift to lower masses when vertical stratification is fully resolved. However, the fundamental qualitative mechanisms discovered here remains physically robust. The dynamical attractor found in this paper corresponds to a morphological feature that should persist even in fully 3D disks. Therefore, while the exact planetary mass threshold for trap failure scales with dimensionality, the existence of dynamical traps in dust-rich environments should be a universal physical mechanism. Given its critical implications for the final architecture of planetary systems, quantifying this exact bifurcation mass via fully stratified, self-consistent 3D hydrodynamical simulations is a crucial next step for the field.

\subsection{Spatial scope and physical context}

In our framework, we define $r_0 = 3\,\mathrm{au}$ as our reference scale unit. The inner disk ($a_0/r_0 = 1$) corresponds to the vicinity of the water-ice line, while the outer disk ($a_0/r_0 = 10$) denotes the region well beyond the snow line ($r \approx 30\,\mathrm{au}$). Crucially, our simulations show that for embryos initialized at $a_0/r_0 = 1$, lower Stokes numbers ($\mathrm{St} = 0.10$) consistently yield sustained inward migration trajectories, as dust-induced torques remain negative or insufficient to counteract the gas torque. Interior to the ice line ($r < 3\,\mathrm{au}$), collisional fragmentation restricts dry silicates to much smaller values \citep{Batygin2022}, making our assumption of a spatially uniform $\mathrm{St} = 0.10$ a simplification.  Consequently, our main findings on dust-driven trapping and outward migration specifically target the outer icy disk domain ($r \ge 3\,\mathrm{au}$), where moderately coupled icy pebbles ($\mathrm{St} = 0.10\text{--}0.55$) generate robust dust asymmetries capable of substantially modifying planetary migration.

\subsection{Evolution of solids fraction}

It is worth emphasizing that the local dust-to-gas mass ratio, $\epsilon$, in the vicinity of a forming planet is not a static property throughout the disk lifetime, but rather evolves dynamically across different stages of planet formation \citep[e.g.,][]{Chiang2010, Bergin_2017}. While large-scale ALMA surveys demonstrate that disk-averaged gas-to-dust mass ratios evolve over million-year timescales and systematically drop below canonical ISM values \citep[e.g.,][]{Ansdell_2016, Bergin_2017}, these global measurements cannot resolve the fine-scale variability occurring near an embedded embryo. On local scales, the conversion of pebbles into macroscopic planetesimals and planetary cores depletes the surrounding dust fluid, creating local dips in $\epsilon$, whereas aerodynamic drift pileups, pressure trapping, and debris generated by early protoplanetary collisions can produce prominent local spikes in the solid fraction \citep[e.g.,][]{Chiang2010, Savvidou2025}. Consequently, the values of $\epsilon$ adopted in hydrodynamic simulations represent illustrative snapshots of local solid availability rather than temporally uniform constants.

\subsection{Astrophysical implications}

Our parameter space exploration shows that dust-induced torques can establish stable equilibrium regions depending on the physical properties of the disk and planetary masses. In high-metallicity environments, we observe trapping not only at near 15 au, but also the emergence of an outer confinement zone near 60 au. Furthermore, the existence of a transition region—where the migration is inward or outward fundamentally depends on the initial orbital position—indicates that planetary survival is not universally guaranteed by the global structure of the disk alone. In marginal regimes, an embryo formed in the outer disk may become successfully trapped, while an identical embryo formed further inward undergoes monotonic decay. This observation suggests that the ultimate survival and final architecture of a planetary system are deeply tied to the embryos' specific formation history and initial birth locations.
The efficiency of this outer trapping mechanism is strongly modulated by the particle size that exert torque or aerodynamic coupling. A Stokes number of ${\rm St} \sim 0.5$ corresponds physically to centimeter-sized particles located at large orbital distances, typically between 20 and 50 au for our gas surface density of $\Sigma_0 = 6.37 \times 10^{-4}$ ($\Sigma_{\text{gas}} \approx 630\,\mathrm{g\,cm^{-2}}$ at $3\,\mathrm{au}$), where the background gas surface density is intrinsically lower. The emergence of stable confinement zones at these large distances provides a compelling theoretical framework for the demographic population of exoplanets in distant orbits. By efficiently halting inward migration, dust-modulated traps offer a robust mechanism for low-mass planetary cores to form and survive well beyond the inner disk, preserving the architecture of wide-orbit planetary systems prior to disk dissipation.
Crucially, the dynamic preservation of these wide-orbit traps provides a direct theoretical link to the submillimeter rings and asymmetric substructures frequently resolved by high-resolution interferometric surveys—such as the DSHARP \citep{Andrews2018} and ODISEA \citep{Cieza2021, Orcajo2025} programs—as well as the observed architecture of giant planets at wide orbits \citep[e.g.,][]{Marois2008, Bowler2016}. The physical connection between these disk substructures and forming planets has been heavily reinforced by recent direct-imaging discoveries, most notably in the PDS~70 and WISPIT~2 systems, where young planets are caught in the act of carving out prominent gas and dust cavities \citep{Keppler2018, Close2025}. Within this observational framework, the dust-mediated outward migration identified in our simulations offers a highly compelling formation pathway: a planetary core undergoing rapid outward relocation before being permanently locked at an outer dynamical attractor can efficiently remain in a solid-rich feeding zone to accumulate a massive envelope, acting as the primary seed for wide-orbit giants.

\subsection{Limitations and Future Directions}

It is important to acknowledge the limitations of our numerical exploration, particularly regarding the evolution of orbital eccentricity. In the majority of the oscillatory regimes, the planetary eccentricity remains bounded. However, the continuous, asymmetric gravitational pull from the dense leading dust filament acts as a periodic forcing. For specific parameters—particularly for planets approaching the critical mass threshold—this interaction induces a gradual secular growth in eccentricity. Once the eccentricity surpasses a critical threshold ($e \gtrsim 0.2$), the planet's radial excursions become large, making our confined comoving computational grid insufficient to resolve the spiral wake. Studying the evolution and ultimate fate of these highly eccentric cases requires improvements of the used domain or reference radius, such as replacing the semi-major axis $a_{\rm p}$ with the instantaneous orbital radius $r_{\rm p}$ in the comoving transformation.
Furthermore, conducting long-term, multi-fluid hydrodynamic integrations to fully map the phase-space attractors is numerically challenging. However, the robust, quasi-periodic nature of the trajectories within these traps suggests that the complex dynamics can be abstracted into a simplified theoretical framework. Because the confinement mechanism fundamentally relies on a temporal mismatch between the planet's migration velocity and the restructuring timescale of the co-orbital dust, this system could be effectively modeled using a non-linear driven oscillator.

\section{Summary and Conclusions}
\label{sec:summary}

We performed 144 two-fluid hydrodynamical simulations using \texttt{FARGO3D} in a comoving reference frame to map the self-consistent orbital evolution of low-mass planets across a six-dimensional parameter space ($a_0$, $M_p$, ${\rm St}$, $\epsilon$, $\alpha$, $\Sigma_0$). Our findings underscore that allowing for the self-consistent motion of the planet is indispensable for capturing the true long-term dynamical fate of planetary embryos in dust-rich environments.
Our main conclusions are:

\paragraph{Emergence of dynamical attractors} Dust traps do not merely provide instantaneous positive torques. They behave as robust dynamical attractors. Low-mass planets ($M_{p} \le 4 M_{\oplus}$) migrating from both the inner and outer disk can converge to the same radial confinement zones, exhibiting stable oscillatory migration. High-metallicity disks can establish multiple equilibrium regions (e.g., near 5 au and 20 au), demonstrating that trapping is globally viable, although planetary survival in transition regimes remains highly sensitive to the embryo's initial birth location.
    
\paragraph{Self-induced confinement} These orbital traps are dynamically self-induced and regulated by the continuous momentum exchange between the planet, the co-orbital solid material and the gaseous torque. This establishes that robust orbital confinement and sustained positive torques can be achieved entirely through local dust-planet interactions, without the need to invoke pre-existing mechanisms, like pressure maxima or dead-zones in the background gas disk. The survival of the trap is driven by a continuous dynamical pursuit and the periodic hydrodynamic restoration of the leading dust filament and trailing dust cavity.

\paragraph{Deterministic evolution over stochastic migration} High local dust concentrations ($\epsilon=0.05$) trigger a robust dust-gas feedback loop that allows for the existence of positive torques. This active feedback counteracts the smoothing effects of turbulent diffusion, preserving asymmetries even in highly viscous environments. While the lowest-mass embryos may exhibit slightly noisy confinement, the evolution for super-Earths is bounded and deterministic, challenging previous results that dust-driven migration inevitably degrades into a chaotic random walk.

\paragraph{Dust-mediated runaway migration and dimensionality} A critical mass bifurcation dictates the planetary fate. Cores reaching $M_{p} \approx 10 M_{\oplus}$ possess enough mass to overcome the trap, breaking the confinement and entering into a runaway migration regime. This demonstrates that in 2D frameworks, runaway migration becomes dominant at significantly higher planetary masses than those recently predicted by fully stratified 3D models.

\paragraph{Universal scaling} The dust-modulated trapping mechanism operates under a universal physical scaling. By filtering out absolute spatial and temporal coordinates, we reveal that planetary trajectories universally collapse into identical, self-similar limit cycles in phase space. This demonstrates that the confinement is fundamentally dictated by local hydrodynamic conditions and is entirely independent of the planet's prior orbital history.

\paragraph{Spatial applicability and dynamic solid environments} The viability of dust-driven trapping is intimately linked to local disk conditions. This mechanism operates most efficiently in the outer disk (beyond the water-ice line, $r \ge 3$ au), where moderately coupled icy pebbles are expected. Furthermore, because the local dust-to-gas mass ratio evolves dynamically, these orbital attractors are highly favored in regions with enhanced solid concentrations, such as those produced by aerodynamic drift or debris generation following early planetary collisions.

\vspace{1em}

Ultimately, the self-consistent simulations presented here demonstrate that dust-induced confinement zones are not transient disk features, but robust dynamical attractors. However, the broader implication of our work extends beyond the confirmation of the existence of these traps. By allowing the planet to dynamically respond to its own  feedback in the gas/dust disk, we uncover a remarkably diverse spectrum of migratory solutions—from the deterministic trapping of super-Earths to the rapid inward/outward migration of planetary cores. This wide-range of possible solutions is inaccessible to classical fixed-orbit approximations. These diverse outcomes clearly demonstrate that planetary motion can not be neglected or decoupled when predicting orbital evolution.

\begin{acknowledgments}
We thank the reviewer, Matthew Clement, for his comments and constructive suggestions, which significantly improved the quality and clarity of this manuscript. The authors acknowledge the computational resources provided by the High-Performance Computing infrastructure of the Faculty of Engineering and Sciences at Universidad Adolfo Ibáñez, which were instrumental in generating the core results presented in this work. P.~B.~L. acknowledges  support from ANID, QUIMAL fund ASTRO21-0039 and FONDECYT project 1231205. LG and PBL acknowledge the support of the Predoctoral Fellowship program from the Directorate of Research and Academic Postgraduate Studies, Faculty of Engineering and Sciences at Universidad Adolfo Ibáñez. J.~L.~G. and O.~M.~G. are partially supported by PIP-2971 from CONICET (Argentina) and by PID-SG004 from UNLP (Argentina).
\end{acknowledgments}

\software{\texttt{FARGO3D} \citep{Benitez-Llambay2016a, Benitez-Llambay2019, Benitez-Llambay2023, Ramos2025}, Matplotlib \citep{Hunter:2007}}.
          
\appendix

\section{Invariance of dust diffusion in comoving coordinates}
\label{apendix:equations}

The dust continuity equation that accounts for the diffusion flux is
\begin{equation}
\frac{\partial \Sigma_{\rm dust}}{\partial t} + \nabla\!\cdot\left(\Sigma_{\rm dust} \mathbf{v}_{\rm d} + \mathbf{j}_{\rm d}\right) = 0\,, \label{eq:cont_phys}
\end{equation}
where the diffusive flux $\mathbf{j}_{\rm d}$ is
\begin{equation}
\mathbf{j}_{\rm d} = -D_{\rm d}\,\left(\Sigma_{\rm gas}+\Sigma_{\rm dust}\right)\,\nabla\!\left(\frac{\Sigma_{\rm dust}}{\Sigma_{\rm gas}}\right)\,. 
\label{eq:flux_phys}
\end{equation}
After applying the comoving transformation of \cite{Ramos2025} and defining the comoving surface density and comoving diffusion flux as
\begin{equation}
\Sigma'_{\rm gas/dust} = a_{\rm p}^{2}\Sigma_{\rm gas/dust}\,,
\end{equation}
\begin{equation}
\mathbf{j}'_{\rm d} = -D'_{\rm d}\,(\Sigma'_{\rm gas}+\Sigma'_{\rm dust})\,\nabla'\!\left(\frac{\Sigma'_{\rm dust}}{\Sigma'_{\rm gas}}\right)\,,
\end{equation}
with $D_{\rm d} = D'_{\rm d} D_{\rm p}$, we obtain
\begin{equation}
\frac{\partial \Sigma'_{\rm dust}}{\partial t'} + \nabla'\!\cdot\left(\Sigma'_{\rm dust} \mathbf{u'}_{\rm d} + \frac{D_{\rm p}}{\Omega_{\rm p} a_{\rm p}^2} \mathbf{j}'_{\rm d}\right) = 0\,.
\end{equation}
We assume a dust diffusion model that scales as the turbulent viscosity of the $\alpha$-disk model, such that $\nu \alpha_v c_{\rm s} h r \propto a_{\rm p}^2 \Omega_{\rm p}$. Because the diffusion coefficient inherits this identical scaling (see section \ref{sec:numerical_method}), the prefactor $D_{\rm p}/(\Omega_{\rm p} a_{\rm p}^2)$ reduces to unity. Therefore, the resulting continuity equation becomes
\begin{equation}
\frac{\partial \Sigma'_{\rm dust}}{\partial t'} + \nabla'\!\cdot\left(\Sigma'_{\rm dust} \mathbf{u'}_{\rm d} + \mathbf{j}'_{\rm d}\right) = 0\,,
\end{equation}
which is strictly invariant under the comoving transformation. 

\section{Summary of simulation runs}
\label{apendix:table}
In Table\,\ref{tab:fullresult} we present result for our full parameter exploration. 

\startlongtable
\begin{deluxetable}{l l c c c c c r l c}
\tabletypesize{\small}
\tablecaption{\linespread{0.8}\selectfont Summary of our simulations. Input parameters include initial semi-major axis ($a_0$), planet mass ($M_{\rm p}$), Stokes number (${\rm St}$), disk metallicity ($\epsilon$), turbulent viscosity ($\alpha$), and initial surface density ($\Sigma_0$). Outcomes indicate the final Migration Regime (Inward, Outward, or Trapped), final semi-major axis ($a_f$), and eccentricity ($e_f$). For Trapped cases, $a_{\rm f}$ and $e_{\rm f}$ are time-averaged over the last 240 orbits rather than instantaneous. Gray rows indicate reference cases converging into the dynamic trap (Fig.\,\ref{fig:fig5}). Asterisks ($^*$) denote integrations halted due to the planet's eccentricity exceeding the computational grid boundaries. Daggers ($^\dagger$) highlight the 28 representative simulations plotted in Fig.\,\ref{fig:fig4}.}

\renewcommand{\arraystretch}{0.4}
\setlength{\tabcolsep}{3pt}

\tablehead{
    Sim. & \colhead{$a_0 (r_0)$} & \colhead{$M_{\rm p}\, (M_\oplus)$}& \colhead{${\rm St}$}& \colhead{$\epsilon$} & \colhead{$\alpha$} & \colhead{$\Sigma_0\, (M_0/r_0^2)$} & \colhead{$a_{f}\, (r_0)$}& \colhead{$e_{f}$} & \colhead{Mig. Reg.} 
}
\startdata
S1 & 1 & 2.67 & 0.10 & 0.01 & $10^{-4}$ & $6.37\times 10^{-4}$ & 0.97 & $2.40\times 10^{-5}$ &Inward \\
S2 & 1 & 2.67 & 0.10 & 0.01 & $10^{-4}$ & $5.00\times 10^{-3}$ & 0.90 & $9.10\times 10^{-5}$ &Inward \\
S3 & 1 & 2.67 & 0.10 & 0.01 & $10^{-3}$ & $6.37\times 10^{-4}$ & 0.98 & $2.00\times 10^{-5}$ &Inward \\
S4 & 1 & 2.67 & 0.10 & 0.01 & $10^{-3}$ & $5.00\times 10^{-3}$ & 0.88 & $9.10\times 10^{-5}$ &Inward \\
S5 & 1 & 2.67 & 0.10 & 0.05 & $10^{-4}$ & $6.37\times 10^{-4}$ & 0.97 & $2.50\times 10^{-5}$ &Inward \\
S6 & 1 & 2.67 & 0.10 & 0.05 & $10^{-4}$ & $5.00\times 10^{-3}$ & 4.64 & $1.48\times 10^{-3}$ &Trapped \\
S7 & 1 & 2.67 & 0.10 & 0.05 & $10^{-3}$ & $6.37\times 10^{-4}$ & 0.99 & $2.00\times 10^{-5}$ &Inward \\
S8 & 1 & 2.67 & 0.10 & 0.05 & $10^{-3}$ & $5.00\times 10^{-3}$ & 0.90 & $9.50\times 10^{-5}$ &Inward \\
S9  & 1 & 2.67 & 0.26 & 0.01 & $10^{-4}$ & $6.37\times 10^{-4}$ & 0.97 & $1.70\times 10^{-5}$ &Inward \\
S10 & 1 & 2.67 & 0.26 & 0.01 & $10^{-4}$ & $5.00\times 10^{-3}$ & 0.90 & $6.00\times 10^{-5}$ &Inward \\
S11 & 1 & 2.67 & 0.26 & 0.01 & $10^{-3}$ & $6.37\times 10^{-4}$ & 0.99 & $1.80\times 10^{-5}$ &Inward \\
S12 & 1 & 2.67 & 0.26 & 0.01 & $10^{-3}$ & $5.00\times 10^{-3}$ & 0.92 & $8.40\times 10^{-5}$ &Inward \\
S13 & 1 & 2.67 & 0.26 & 0.05 & $10^{-4}$ & $6.37\times 10^{-4}$ & 1.19 &$1.57\times 10^{-4}$  &Outward \\
S14 & 1 & 2.67 & 0.26 & 0.05 & $10^{-4}$ & $5.00\times 10^{-3}$ & 2.55 &$1.56\times 10^{-3}$  &Trapped \\
S15 & 1 & 2.67 & 0.26 & 0.05 & $10^{-3}$ & $6.37\times 10^{-4}$ & 1.12 &$4.10\times 10^{-5}$  &Outward \\
S16 & 1 & 2.67 & 0.26 & 0.05 & $10^{-3}$ & $5.00\times 10^{-3}$ & 2.77 &$6.40\times 10^{-4}$  &Outward \\
S17 & 1 & 2.67 & 0.55 & 0.01 & $10^{-4}$ & $6.37\times 10^{-4}$ & 0.98 & $1.50\times 10^{-5}$ &Inward \\
S18 & 1 & 2.67 & 0.55 & 0.01 & $10^{-4}$ & $5.00\times 10^{-3}$ & 2.73 & $5.98\times 10^{-4}$ &Outward \\
S19 & 1 & 2.67 & 0.55 & 0.01 & $10^{-3}$ & $6.37\times 10^{-4}$ & 1.01 & $1.70\times 10^{-5}$ &Outward \\
S20$^\dagger$ & 1 & 2.67 & 0.55 & 0.01 & $10^{-3}$ & $5.00\times 10^{-3}$ & 1.06 & $8.70\times 10^{-5}$ &Outward \\
S21 & 1 & 2.67 & 0.55 & 0.05 & $10^{-4}$ & $6.37\times 10^{-4}$ & 1.21 & $1.74\times 10^{-4}$ &Outward \\
S22 & 1 & 2.67 & 0.55 & 0.05 & $10^{-4}$ & $5.00\times 10^{-3}$ & 2.92 & $2.27\times 10^{-3}$ &Trapped \\
S23 & 1 & 2.67 & 0.55 & 0.05 & $10^{-3}$ & $6.37\times 10^{-4}$ & 1.41 & $1.33\times 10^{-4}$ &Outward \\
S24$^\dagger$ & 1 & 2.67 & 0.55 & 0.05 & $10^{-3}$ & $5.00\times 10^{-3}$ & 2.97 & $1.75\times 10^{-3}$ &Trapped \\
S25 & 1 & 4.00 & 0.10 & 0.01 & $10^{-4}$ & $6.37\times 10^{-4}$ & 0.96 & $3.30\times 10^{-5}$ &Inward \\
S26 & 1 & 4.00 & 0.10 & 0.01 & $10^{-4}$ & $5.00\times 10^{-3}$ & 0.87 & $1.14\times 10^{-4}$ &Inward \\
S27 & 1 & 4.00 & 0.10 & 0.01 & $10^{-3}$ & $6.37\times 10^{-4}$ & 0.98 & $2.70\times 10^{-5}$ &Inward \\
S28$^\dagger$ & 1 & 4.00 & 0.10 & 0.01 & $10^{-3}$ & $5.00\times 10^{-3}$ & 0.85 & $1.13\times 10^{-4}$ &Inward \\
S29 & 1 & 4.00 & 0.10 & 0.05 & $10^{-4}$ & $6.37\times 10^{-4}$ & 0.95 & $3.70\times 10^{-5}$ &Inward \\
S30 & 1 & 4.00 & 0.10 & 0.05 & $10^{-4}$ & $5.00\times 10^{-3}$ & 0.82 & $1.30\times 10^{-4}$ &Inward \\
S31 & 1 & 4.00 & 0.10 & 0.05 & $10^{-3}$ & $6.37\times 10^{-4}$ & 0.98 & $2.70\times 10^{-5}$ &Inward \\
S32$^\dagger$ & 1 & 4.00 & 0.10 & 0.05 & $10^{-3}$ & $5.00\times 10^{-3}$ & 0.84 & $1.20\times 10^{-4}$ &Inward \\
S33 & 1 & 4.00 & 0.26 & 0.01 & $10^{-4}$ & $6.37\times 10^{-4}$ & 0.97 & $2.70\times 10^{-5}$ &Inward \\
S34 & 1 & 4.00 & 0.26 & 0.01 & $10^{-4}$ & $5.00\times 10^{-3}$ & 0.90 & $9.40\times 10^{-5}$ &Inward \\
S35 & 1 & 4.00 & 0.26 & 0.01 & $10^{-3}$ & $6.37\times 10^{-4}$ & 0.98 & $2.50\times 10^{-5}$ &Inward \\
S36$^\dagger$ & 1 & 4.00 & 0.26 & 0.01 & $10^{-3}$ & $5.00\times 10^{-3}$ & 0.89 & $1.04\times 10^{-4}$ &Inward \\
S37 & 1 & 4.00 & 0.26 & 0.05 & $10^{-4}$ & $6.37\times 10^{-4}$ & 1.18 & $1.90\times 10^{-4}$ &Outward \\
S38 & 1 & 4.00 & 0.26 & 0.05 & $10^{-4}$ & $5.00\times 10^{-3}$ & 2.97 & $2.15\times 10^{-3}$ &Trapped \\
S39 & 1 & 4.00 & 0.26 & 0.05 & $10^{-3}$ & $6.37\times 10^{-4}$ & 1.11 & $4.20\times 10^{-5}$ &Outward \\
S40$^\dagger$ & 1 & 4.00 & 0.26 & 0.05 & $10^{-3}$ & $5.00\times 10^{-3}$ & 6.28 & $3.06\times 10^{-3}$ &Trapped \\
S41 & 1 & 4.00 & 0.55 & 0.01 & $10^{-4}$ & $6.37\times 10^{-4}$ & 0.98 & $2.30\times 10^{-5}$ &Inward \\
S42$^\dagger$ & 1 & 4.00 & 0.55 & 0.01 & $10^{-4}$ & $5.00\times 10^{-3}$ & 0.93 & $6.80\times 10^{-5}$ &Inward \\
S43$^\dagger$ & 1 & 4.00 & 0.55 & 0.01 & $10^{-3}$ & $6.37\times 10^{-4}$ & 1.00 & $2.20\times 10^{-5}$ &Inward \\
S44$^\dagger$ & 1 & 4.00 & 0.55 & 0.01 & $10^{-3}$ & $5.00\times 10^{-3}$ & 1.01 & $9.40\times 10^{-5}$ &Outward \\
S45 & 1 & 4.00 & 0.55 & 0.05 & $10^{-4}$ & $6.37\times 10^{-4}$ & 1.23 & $2.32\times 10^{-4}$ &Outward \\
S46$^\dagger$ & 1 & 4.00 & 0.55 & 0.05 & $10^{-4}$ & $5.00\times 10^{-3}$ & 3.35 & $2.58\times 10^{-3}$ &Trapped \\
S47$^\dagger$ & 1 & 4.00 & 0.55 & 0.05 & $10^{-3}$ & $6.37\times 10^{-4}$ & 1.43 & $1.58\times 10^{-4}$ &Outward \\
\rowcolor{gray!30} S48$^\dagger$ & 1 & 4.00 & 0.55 & 0.05 & $10^{-3}$ & $5.00\times 10^{-3}$ & 5.16 & $4.04\times 10^{-3}$ &Trapped \\
S49 & 1 & 10.0 & 0.10 & 0.01 & $10^{-4}$ & $6.37\times 10^{-4}$ & 0.91 & $7.10\times 10^{-5}$ &Inward \\
S50 & 1 & 10.0 & 0.10 & 0.01 & $10^{-4}$ & $5.00\times 10^{-3}$ & 0.78 & $1.81\times 10^{-4}$ &Inward \\
S51 & 1 & 10.0 & 0.10 & 0.01 & $10^{-3}$ & $6.37\times 10^{-4}$ & 0.94 & $5.60\times 10^{-5}$ &Inward \\
S52 & 1 & 10.0 & 0.10 & 0.01 & $10^{-3}$ & $5.00\times 10^{-3}$ & 0.73 & $1.81\times 10^{-4}$ &Inward \\
S53 & 1 & 10.0 & 0.10 & 0.05 & $10^{-4}$ & $6.37\times 10^{-4}$ & 0.89 & $7.10\times 10^{-5}$ &Inward \\
S54 & 1 & 10.0 & 0.10 & 0.05 & $10^{-4}$ & $5.00\times 10^{-3}$ & 0.68 & $2.23\times 10^{-4}$ &Inward \\
S55 & 1 & 10.0 & 0.10 & 0.05 & $10^{-3}$ & $6.37\times 10^{-4}$ & 0.92 & $5.80\times 10^{-5}$ &Inward \\
S56 & 1 & 10.0 & 0.10 & 0.05 & $10^{-3}$ & $5.00\times 10^{-3}$ & 0.67 & $2.06\times 10^{-4}$ &Inward \\
S57 & 1 & 10.0 & 0.26 & 0.01 & $10^{-4}$ & $6.37\times 10^{-4}$ & 0.91 & $6.70\times 10^{-5}$ &Inward \\
S58 & 1 & 10.0 & 0.26 & 0.01 & $10^{-4}$ & $5.00\times 10^{-3}$ & 0.81 & $1.54\times 10^{-4}$ &Inward \\
S59 & 1 & 10.0 & 0.26 & 0.01 & $10^{-3}$ & $6.37\times 10^{-4}$ & 0.94 & $5.30\times 10^{-5}$ &Inward \\
S60 & 1 & 10.0 & 0.26 & 0.01 & $10^{-3}$ & $5.00\times 10^{-3}$ & 0.76 & $1.73\times 10^{-4}$ &Inward \\
S61 & 1 & 10.0 & 0.26 & 0.05 & $10^{-4}$ & $6.37\times 10^{-4}$ & 0.93 & $9.00\times 10^{-5}$ &Inward \\
S62 & 1 & 10.0 & 0.26 & 0.05 & $10^{-4}$ & $5.00\times 10^{-3}$ & 2.00 & $1.22\times 10^{-3}$ &Trapped \\
S63 & 1 & 10.0 & 0.26 & 0.05 & $10^{-3}$ & $6.37\times 10^{-4}$ & 0.96 & $4.70\times 10^{-5}$ &Inward \\
S64 & 1 & 10.0 & 0.26 & 0.05 & $10^{-3}$ & $5.00\times 10^{-3}$ & 0.73 & $1.79\times 10^{-4}$ &Inward \\
S65 & 1 & 10.0 & 0.55 & 0.01 & $10^{-4}$ & $6.37\times 10^{-4}$ & 0.93 & $5.50\times 10^{-5}$ &Inward \\
S66 & 1 & 10.0 & 0.55 & 0.01 & $10^{-4}$ & $5.00\times 10^{-3}$ & 0.85 & $1.10\times 10^{-4}$ &Inward \\
S67 & 1 & 10.0 & 0.55 & 0.01 & $10^{-3}$ & $6.37\times 10^{-4}$ & 0.96 & $4.80\times 10^{-5}$ &Inward \\
S68$^\dagger$ & 1 & 10.0 & 0.55 & 0.01 & $10^{-3}$ & $5.00\times 10^{-3}$ & 0.80 & $1.52\times 10^{-4}$ &Inward \\
S69 & 1 & 10.0 & 0.55 & 0.05 & $10^{-4}$ & $6.37\times 10^{-4}$ & 0.98 & $3.40\times 10^{-5}$ &Inward \\
S70 & 1 & 10.0 & 0.55 & 0.05 & $10^{-4}$ & $5.00\times 10^{-3}$ & 2.71 & $2.01\times 10^{-3}$ &Trapped \\
S71 & 1 & 10.0 & 0.55 & 0.05 & $10^{-3}$ & $6.37\times 10^{-4}$ & 1.03 & $3.40\times 10^{-4}$ &Outward \\
S72$^\dagger$ & 1 & 10.0 & 0.55 & 0.05 & $10^{-3}$ & $5.00\times 10^{-3}$ & 3.82 & $2.62\times 10^{-3}$ &Trapped \\
S73 & 10 & 2.67 & 0.10 & 0.01 & $10^{-4}$ & $6.37\times 10^{-4}$ & 8.52 & $2.66\times 10^{-4}$ &Inward \\
S74 & 10 & 2.67 & 0.10 & 0.01 & $10^{-4}$ & $5.00\times 10^{-3}$ & 5.47 & $9.92\times 10^{-4}$ &Inward \\
S75 & 10 & 2.67 & 0.10 & 0.01 & $10^{-3}$ & $6.37\times 10^{-4}$ & 7.18 & $2.31\times 10^{-4}$ &Inward \\
S76 & 10 & 2.67 & 0.10 & 0.01 & $10^{-3}$ & $5.00\times 10^{-3}$ & 4.06 & $6.16\times 10^{-4}$ &Inward \\
S77 & 10 & 2.67 & 0.10 & 0.05 & $10^{-4}$ & $6.37\times 10^{-4}$ & 17.61 & $2.39\times 10^{-3}$ &Outward \\
S78 & 10 & 2.67 & 0.10 & 0.05 & $10^{-4}$ & $5.00\times 10^{-3}$ & 4.03 & $5.83\times 10^{-4}$ &Trapped \\
S79 & 10 & 2.67 & 0.10 & 0.05 & $10^{-3}$ & $6.37\times 10^{-4}$ & 7.33 & $2.53\times 10^{-4}$ &Inward \\
S80 & 10 & 2.67 & 0.10 & 0.05 & $10^{-3}$ & $5.00\times 10^{-3}$ & 3.84 & $6.24\times 10^{-4}$ &Inward \\
S81 & 10 & 2.67 & 0.26 & 0.01 & $10^{-4}$ & $6.37\times 10^{-4}$ & 20.82 & $1.01\times 10^{-3}$ &Outward \\
S82 & 10 & 2.67 & 0.26 & 0.01 & $10^{-4}$ & $5.00\times 10^{-3}$ & 7.10 & $1.28\times 10^{-3}$ &Inward \\
S83 & 10 & 2.67 & 0.26 & 0.01 & $10^{-3}$ & $6.37\times 10^{-4}$ & 7.97 & $2.39\times 10^{-4}$ &Inward \\
S84 & 10 & 2.67 & 0.26 & 0.01 & $10^{-3}$ & $5.00\times 10^{-3}$ & 7.31 & $1.32\times 10^{-3}$ &Inward \\
S85 & 10 & 2.67 & 0.26 & 0.05 & $10^{-4}$ & $6.37\times 10^{-4}$ & 14.61 & $2.53\times 10^{-3}$ &Trapped \\
S86 & 10 & 2.67 & 0.26 & 0.05 & $10^{-4}$ & $5.00\times 10^{-3}$ & 4.13  & $3.09\times 10^{-3}$ &Trapped \\
S87 & 10 & 2.67 & 0.26 & 0.05 & $10^{-3}$ & $6.37\times 10^{-4}$ & 21.54 & $2.13\times 10^{-3}$ &Trapped \\
S88 & 10 & 2.67 & 0.26 & 0.05 & $10^{-3}$ & $5.00\times 10^{-3}$ & 6.02  & $2.92\times 10^{-3}$ &Trapped \\
S89 & 10 & 2.67 & 0.55 & 0.01 & $10^{-4}$ & $6.37\times 10^{-4}$ & 26.48 & $1.38\times 10^{-3}$ &Outward \\
S90 & 10 & 2.67 & 0.55 & 0.01 & $10^{-4}$ & $5.00\times 10^{-3}$ & 7.34  & $1.79\times 10^{-3}$ &Trapped \\
S91 & 10 & 2.67 & 0.55 & 0.01 & $10^{-3}$ & $6.37\times 10^{-4}$ & 15.12 & $6.43\times 10^{-4}$ &Outward \\
S92$^\dagger$ & 10 & 2.67 & 0.55 & 0.01 & $10^{-3}$ & $5.00\times 10^{-3}$ & 9.60 & $2.40\times 10^{-3}$ &Trapped \\
S93 & 10 & 2.67 & 0.55 & 0.05 & $10^{-4}$ & $6.37\times 10^{-4}$ & 16.25 & $4.45\times 10^{-3}$ &Trapped \\
S94 & 10 & 2.67 & 0.55 & 0.05 & $10^{-4}$ & $5.00\times 10^{-3}$ & 4.25 & $4.18\times 10^{-3}$ &Trapped \\
S95 & 10 & 2.67 & 0.55 & 0.05 & $10^{-3}$ & $6.37\times 10^{-4}$ & 19.06 & $3.57\times 10^{-3}$ &Trapped \\
S96$^\dagger$ & 10 & 2.67 & 0.55 & 0.05 & $10^{-3}$ & $5.00\times 10^{-3}$ & 4.82 & $3.64\times 10^{-3}$ &Trapped \\
S97 & 10 & 4.00 & 0.10 & 0.01 & $10^{-4}$ & $6.37\times 10^{-4}$ & 8.28 & $3.31\times 10^{-4}$ &Inward \\
S98 & 10 & 4.00 & 0.10 & 0.01 & $10^{-4}$ & $5.00\times 10^{-3}$ & 5.47 & $1.31\times 10^{-3}$ &Inward \\
S99 & 10 & 4.00 & 0.10 & 0.01 & $10^{-3}$ & $6.37\times 10^{-4}$ & 6.79 & $2.73\times 10^{-4}$ &Inward \\
S100$^\dagger$& 10 & 4.00 & 0.10 & 0.01 & $10^{-3}$ & $5.00\times 10^{-3}$ & 4.05 & $7.75\times 10^{-4}$ &Inward \\
S101 & 10 & 4.00 & 0.10 & 0.05 & $10^{-4}$ & $6.37\times 10^{-4}$ & 15.97 & $9.13\times 10^{-4}$ &Outward \\
S102 & 10 & 4.00 & 0.10 & 0.05 & $10^{-4}$ & $5.00\times 10^{-3}$ & 5.02 & $1.63\times 10^{-3}$ &Trapped \\
S103 & 10 & 4.00 & 0.10 & 0.05 & $10^{-3}$ & $6.37\times 10^{-4}$ & 6.62 & $2.91\times 10^{-4}$ &Inward \\
S104$^\dagger$ & 10 & 4.00 & 0.10 & 0.05 & $10^{-3}$ & $5.00\times 10^{-3}$ & 3.66 & $7.27\times 10^{-4}$ &Inward \\
S105 & 10 & 4.00 & 0.26 & 0.01 & $10^{-4}$ & $6.37\times 10^{-4}$ & 8.81 & $2.94\times 10^{-4}$ &Inward \\
S106 & 10 & 4.00 & 0.26 & 0.01 & $10^{-4}$ & $5.00\times 10^{-3}$ & 6.53 & $1.29\times 10^{-3}$ &Inward \\
S107 & 10 & 4.00 & 0.26 & 0.01 & $10^{-3}$ & $6.37\times 10^{-4}$ & 7.64 & $2.82\times 10^{-4}$ &Inward \\
S108*$^{,\dagger}$ & 10 & 4.00 & 0.26 & 0.01 & $10^{-3}$ & $5.00\times 10^{-3}$ & -- & -- &Outward \\
S109 & 10 & 4.00 & 0.26 & 0.05 & $10^{-4}$ & $6.37\times 10^{-4}$ & 16.49 & $2.67\times 10^{-3}$ &Trapped \\
S110 & 10 & 4.00 & 0.26 & 0.05 & $10^{-4}$ & $5.00\times 10^{-3}$ & 4.37  & $3.07\times 10^{-3}$ &Trapped \\
S111 & 10 & 4.00 & 0.26 & 0.05 & $10^{-3}$ & $6.37\times 10^{-4}$ & 25.19 & $3.38\times 10^{-3}$ &Trapped \\
S112$^\dagger$ & 10 & 4.00 & 0.26 & 0.05 & $10^{-3}$ & $5.00\times 10^{-3}$ & 6.44 & $3.58\times 10^{-3}$ &Trapped \\
S113 & 10 & 4.00 & 0.55 & 0.01 & $10^{-4}$ & $6.37\times 10^{-4}$ & 27.51 & $1.98\times 10^{-3}$ &Trapped \\
S114$^\dagger$ & 10 & 4.00 & 0.55 & 0.01 & $10^{-4}$ & $5.00\times 10^{-3}$ & 6.98 & $1.84\times 10^{-3}$ &Trapped \\
S115$^\dagger$ & 10 & 4.00 & 0.55 & 0.01 & $10^{-3}$ & $6.37\times 10^{-4}$ & 18.56 & $1.14\times 10^{-3}$ &Outward \\
S116$^\dagger$ & 10 & 4.00 & 0.55 & 0.01 & $10^{-3}$ & $5.00\times 10^{-3}$ & 12.06 & $3.89\times 10^{-3}$ &Trapped \\
S117 & 10 & 4.00 & 0.55 & 0.05 & $10^{-4}$ & $6.37\times 10^{-4}$ & 16.09 & $3.14\times 10^{-3}$ &Trapped \\
S118$^\dagger$ & 10 & 4.00 & 0.55 & 0.05 & $10^{-4}$ & $5.00\times 10^{-3}$ & 4.31 & $3.29\times 10^{-3}$ &Trapped \\
\rowcolor{gray!30} S119$^\dagger$ & 10 & 4.00 & 0.55 & 0.05 & $10^{-3}$ & $6.37\times 10^{-4}$ & 20.41 & $3.94\times 10^{-3}$ &Trapped \\
\rowcolor{gray!30} S120$^\dagger$ & 10 & 4.00 & 0.55 & 0.05 & $10^{-3}$ & $5.00\times 10^{-3}$ & 5.16 & $4.06\times 10^{-3}$ &Trapped \\
S121 & 10 & 10.0 & 0.10 & 0.01 & $10^{-4}$ & $6.37\times 10^{-4}$ & 7.14 & $4.11\times 10^{-4}$ &Inward \\
S122 & 10 & 10.0 & 0.10 & 0.01 & $10^{-4}$ & $5.00\times 10^{-3}$ & 5.77 & $3.70\times 10^{-3}$ &Inward \\
S123 & 10 & 10.0 & 0.10 & 0.01 & $10^{-3}$ & $6.37\times 10^{-4}$ & 5.68 & $3.77\times 10^{-4}$ &Inward \\
S124 & 10 & 10.0 & 0.10 & 0.01 & $10^{-3}$ & $5.00\times 10^{-3}$ & 5.63 & $2.06\times 10^{-3}$ &Inward \\
S125  & 10 & 10.0 & 0.10 & 0.05 & $10^{-4}$ & $6.37\times 10^{-4}$ & 6.00 & $3.76\times 10^{-4}$ &Inward \\
S126* & 10 & 10.0 & 0.10 & 0.05 & $10^{-4}$ & $5.00\times 10^{-3}$ & -- & -- &Outward \\
S127  & 10 & 10.0 & 0.10 & 0.05 & $10^{-3}$ & $6.37\times 10^{-4}$ & 5.45 & $3.87\times 10^{-4}$ &Inward \\
S128* & 10 & 10.0 & 0.10 & 0.05 & $10^{-3}$ & $5.00\times 10^{-3}$ & -- & -- &Outward \\
S129  & 10 & 10.0 & 0.26 & 0.01 & $10^{-4}$ & $6.37\times 10^{-4}$ & 7.69 & $3.83\times 10^{-4}$ &Inward \\
S130* & 10 & 10.0 & 0.26 & 0.01 & $10^{-4}$ & $5.00\times 10^{-3}$ & -- & -- &Outward \\
S131  & 10 & 10.0 & 0.26 & 0.01 & $10^{-3}$ & $6.37\times 10^{-4}$ & 6.17 & $3.60\times 10^{-4}$ &Inward \\
S132* & 10 & 10.0 & 0.26 & 0.01 & $10^{-3}$ & $5.00\times 10^{-3}$ & -- & -- &Outward \\
S133  & 10 & 10.0 & 0.26 & 0.05 & $10^{-4}$ & $6.37\times 10^{-4}$ & 11.81 & $2.18\times 10^{-3}$ &Trapped \\
S134* & 10 & 10.0 & 0.26 & 0.05 & $10^{-4}$ & $5.00\times 10^{-3}$ & -- & -- &Outward \\
S135  & 10 & 10.0 & 0.26 & 0.05 & $10^{-3}$ & $6.37\times 10^{-4}$ & 17.08 & $1.76\times 10^{-3}$ &Trapped \\ S136* & 10 & 10.0 & 0.26 & 0.05 & $10^{-3}$ & $5.00\times 10^{-3}$ & -- & -- &Outward \\
S137  & 10 & 10.0 & 0.55 & 0.01 & $10^{-4}$ & $6.37\times 10^{-4}$ & 8.01 & $6.56\times 10^{-4}$ &Inward \\ 
S138* & 10 & 10.0 & 0.55 & 0.01 & $10^{-4}$ & $5.00\times 10^{-3}$ & -- & -- &Outward \\
S139  & 10 & 10.0 & 0.55 & 0.01 & $10^{-3}$ & $6.37\times 10^{-4}$ & 6.73 & $3.35\times 10^{-4}$ &Inward \\ 
S140*$^{,\dagger}$ & 10 & 10.0 & 0.55 & 0.01 & $10^{-3}$ & $5.00\times 10^{-3}$ & -- & -- &Outward \\
S141  & 10 & 10.0 & 0.55 & 0.05 & $10^{-4}$ & $6.37\times 10^{-4}$ & 13.28 & $6.08\times 10^{-3}$ &Outward \\
S142* & 10 & 10.0 & 0.55 & 0.05 & $10^{-4}$ & $5.00\times 10^{-3}$ & -- & -- &Outward \\
S143  & 10 & 10.0 & 0.55 & 0.05 & $10^{-3}$ & $6.37\times 10^{-4}$ & 19.98 & $3.75\times 10^{-3}$ &Outward \\ 
S144*$^{,\dagger}$ & 10 & 10.0 & 0.55 & 0.05 & $10^{-3}$ & $5.00\times 10^{-3}$ & -- & -- &Outward \\ 
\enddata
\label{tab:fullresult}
\end{deluxetable}


\begin{thebibliography}{}
\expandafter\ifx\csname natexlab\endcsname\relax\def\natexlab#1{#1}\fi
\providecommand{\url}[1]{\href{#1}{#1}}
\providecommand{\dodoi}[1]{doi:~\href{http://doi.org/#1}{\nolinkurl{#1}}}
\providecommand{\doeprint}[1]{\href{http://ascl.net/#1}{\nolinkurl{http://ascl.net/#1}}}
\providecommand{\doarXiv}[1]{\href{https://arxiv.org/abs/#1}{\nolinkurl{https://arxiv.org/abs/#1}}}

\bibitem[{S.~M. {Andrews} {et~al.}(2018){Andrews}, {Huang}, {P{\'e}rez}, {Isella}, {Dullemond}, {Kurtovic}, {Guzm{\'a}n}, {Carpenter}, {Wilner}, {Zhang}, {Zhu}, {Birnstiel}, {Bai}, {Benisty}, {Hughes}, {{\"O}berg}, \& {Ricci}}]{Andrews2018}
{Andrews}, S.~M., {Huang}, J., {P{\'e}rez}, L.~M., {et~al.} 2018, \bibinfo{title}{{The Disk Substructures at High Angular Resolution Project (DSHARP). I. Motivation, Sample, Calibration, and Overview},} \apjl, 869, L41, \dodoi{10.3847/2041-8213/aaf741}

\bibitem[{M. Ansdell {et~al.}(2016)Ansdell, Williams, Marel, Carpenter, Guidi, Hogerheijde, Mathews, Manara, Miotello, Natta, Oliveira, Tazzari, Testi, van Dishoeck, \& van Terwisga}]{Ansdell_2016}
Ansdell, M., Williams, J.~P., Marel, N. v.~d., {et~al.} 2016, \bibinfo{title}{ALMA SURVEY OF LUPUS PROTOPLANETARY DISKS. I. DUST AND GAS MASSES,} The Astrophysical Journal, 828, 46, \dodoi{10.3847/0004-637X/828/1/46}

\bibitem[{{}{Batygin K.} {\&} {}{Morbidelli A.}(2022){Batygin K.} \& {Morbidelli A.}}]{Batygin2022}
{Batygin K.}, \& {Morbidelli A.} 2022, \bibinfo{title}{Self-consistent model for dust-gas coupling in protoplanetary disks,} A\&A, 666, A19, \dodoi{10.1051/0004-6361/202243196}

\bibitem[{P. {Ben{\'\i}tez-Llambay}(2026){Ben{\'\i}tez-Llambay}}]{Benitez-Llambay2026}
{Ben{\'\i}tez-Llambay}, P. 2026, \bibinfo{title}{{Planet{\textendash}disk interaction and evolution},} in Encyclopedia of Astrophysics, Vol.~1, 250--269, \dodoi{10.1016/B978-0-443-21439-4.00033-X}

\bibitem[{P. {Ben{\'\i}tez-Llambay} {et~al.}(2019){Ben{\'\i}tez-Llambay}, {Krapp}, \& {Pessah}}]{Benitez-Llambay2019}
{Ben{\'\i}tez-Llambay}, P., {Krapp}, L., \& {Pessah}, M.~E. 2019, \bibinfo{title}{{Asymptotically Stable Numerical Method for Multispecies Momentum Transfer: Gas and Multifluid Dust Test Suite and Implementation in FARGO3D},} \apjs, 241, 25, \dodoi{10.3847/1538-4365/ab0a0e}

\bibitem[{P. {Ben{\'\i}tez-Llambay} {et~al.}(2015){Ben{\'\i}tez-Llambay}, {Masset}, {Koenigsberger}, \& {Szul{\'a}gyi}}]{Benitez-Llambay2015}
{Ben{\'\i}tez-Llambay}, P., {Masset}, F., {Koenigsberger}, G., \& {Szul{\'a}gyi}, J. 2015, \bibinfo{title}{{Planet heating prevents inward migration of planetary cores},} \nat, 520, 63, \dodoi{10.1038/nature14277}

\bibitem[{P. {Ben{\'\i}tez-Llambay} {\&} F.~S. {Masset}(2016){Ben{\'\i}tez-Llambay} \& {Masset}}]{Benitez-Llambay2016a}
{Ben{\'\i}tez-Llambay}, P., \& {Masset}, F.~S. 2016, \bibinfo{title}{{FARGO3D: A New GPU-oriented MHD Code},} \apjs, 223, 11, \dodoi{10.3847/0067-0049/223/1/11}

\bibitem[{P. {Ben{\'\i}tez-Llambay} {\&} M.~E. {Pessah}(2018){Ben{\'\i}tez-Llambay} \& {Pessah}}]{Benitez-Llambay2018}
{Ben{\'\i}tez-Llambay}, P., \& {Pessah}, M.~E. 2018, \bibinfo{title}{{Torques Induced by Scattered Pebble-flow in Protoplanetary Disks},} \apjl, 855, L28, \dodoi{10.3847/2041-8213/aab2ae}

\bibitem[{P. {Ben{\'\i}tez-Llambay} {et~al.}(2016){Ben{\'\i}tez-Llambay}, {Ramos}, {Beaug{\'e}}, \& {Masset}}]{Benitez-Llambay2016}
{Ben{\'\i}tez-Llambay}, P., {Ramos}, X.~S., {Beaug{\'e}}, C., \& {Masset}, F.~S. 2016, \bibinfo{title}{{Long-term and Large-scale Hydrodynamical Simulations of Migrating Planets},} \apj, 826, 13, \dodoi{10.3847/0004-637X/826/1/13}

\bibitem[{P. Benítez-Llambay {et~al.}(2023)Benítez-Llambay, Krapp, Ramos, \& Kratter}]{Benitez-Llambay2023}
Benítez-Llambay, P., Krapp, L., Ramos, X.~S., \& Kratter, K.~M. 2023, \bibinfo{title}{RAM: Rapid Advection Algorithm on Arbitrary Meshes,} The Astrophysical Journal, 952, 106, \dodoi{10.3847/1538-4357/acd698}

\bibitem[{E.~A. Bergin {\&} J.~P. Williams(2017)Bergin \& Williams}]{Bergin_2017}
Bergin, E.~A., \& Williams, J.~P. 2017, The Determination of Protoplanetary Disk Masses (Springer International Publishing), 1–37, \dodoi{10.1007/978-3-319-60609-5_1}

\bibitem[{B.~P. Bowler(2016)Bowler}]{Bowler2016}
Bowler, B.~P. 2016, \bibinfo{title}{Imaging Extrasolar Giant Planets,} Publications of the Astronomical Society of the Pacific, 128, 102001, \dodoi{10.1088/1538-3873/128/968/102001}

\bibitem[{R.~O. {Chametla} {et~al.}(2025){Chametla}, {Chrenko}, {Masset}, {D'Angelo}, \& {Nesvorn{\'y}}}]{Chametla2025}
{Chametla}, R.~O., {Chrenko}, O., {Masset}, F.~S., {D'Angelo}, G., \& {Nesvorn{\'y}}, D. 2025, \bibinfo{title}{{Dust-void evolution driven by turbulent dust flux can induce runaway migration of Earth-mass planets},} \aap, 698, A21, \dodoi{10.1051/0004-6361/202451869}

\bibitem[{J.-W. {Chen} {\&} M.-K. {Lin}(2018){Chen} \& {Lin}}]{Chen2018}
{Chen}, J.-W., \& {Lin}, M.-K. 2018, \bibinfo{title}{{Dusty disc-planet interaction with dust-free simulations},} \mnras, 478, 2737, \dodoi{10.1093/mnras/sty1166}

\bibitem[{E. Chiang {\&} A. Youdin(2010)Chiang \& Youdin}]{Chiang2010}
Chiang, E., \& Youdin, A. 2010, \bibinfo{title}{Forming Planetesimals in Solar and Extrasolar Nebulae,} Annual Review of Earth and Planetary Sciences, 38, 493, \dodoi{https://doi.org/10.1146/annurev-earth-040809-152513}

\bibitem[{O. {Chrenko} {et~al.}(2024){Chrenko}, {Chametla}, {Masset}, {Baruteau}, \& {Bro{\v{z}}}}]{Chrenko2024}
{Chrenko}, O., {Chametla}, R.~O., {Masset}, F.~S., {Baruteau}, C., \& {Bro{\v{z}}}, M. 2024, \bibinfo{title}{{Pebble-driven migration of low-mass planets in the 2D regime of pebble accretion},} \aap, 690, A41, \dodoi{10.1051/0004-6361/202450922}

\bibitem[{L.~A. {Cieza} {et~al.}(2021){Cieza}, {Gonz{\'a}lez-Ruilova}, {Hales}, {Pinilla}, {Ru{\'\i}z-Rodr{\'\i}guez}, {Zurlo}, {Casassus}, {P{\'e}rez}, {C{\'a}novas}, {Arce-Tord}, {Flock}, {Kurtovic}, {Marino}, {Nogueira}, {Perez}, {Price}, {Principe}, \& {Williams}}]{Cieza2021}
{Cieza}, L.~A., {Gonz{\'a}lez-Ruilova}, C., {Hales}, A.~S., {et~al.} 2021, \bibinfo{title}{{The Ophiuchus DIsc Survey Employing ALMA (ODISEA) - III. The evolution of substructures in massive discs at 3-5 au resolution},} \mnras, 501, 2934, \dodoi{10.1093/mnras/staa3787}

\bibitem[{L.~M. {Close} {et~al.}(2025){Close}, {van Capelleveen}, {Weible}, {Wagner}, {Haffert}, {Males}, {Ilyin}, {Kenworthy}, {Li}, {Long}, {Ertel}, {Ginski}, {Weinberger}, {Follette}, {Liberman}, {Twitchell}, {Johnson}, {Kueny}, {Apai}, {Doyon}, {Foster}, {Gasho}, {Van Gorkom}, {Guyon}, {Kautz}, {McLeod}, {McEwen}, {Pearce}, {Schatz}, {Hedglen}, {Wu}, {Isbell}, {Power}, {Carlson}, {Close}, {Tonucci}, \& {Mars}}]{Close2025}
{Close}, L.~M., {van Capelleveen}, R.~F., {Weible}, G., {et~al.} 2025, \bibinfo{title}{{Wide Separation Planets in Time (WISPIT): Discovery of a Gap H{\ensuremath{\alpha}} Protoplanet WISPIT 2b with MagAO-X},} \apjl, 990, L9, \dodoi{10.3847/2041-8213/adf7a5}

\bibitem[{A. {Crida} {et~al.}(2009){Crida}, {Baruteau}, {Kley}, \& {Masset}}]{Crida2009}
{Crida}, A., {Baruteau}, C., {Kley}, W., \& {Masset}, F. 2009, \bibinfo{title}{{The dynamical role of the circumplanetary disc in planetary migration},} \aap, 502, 679, \dodoi{10.1051/0004-6361/200811608}

\bibitem[{O.~M. {Guilera} {et~al.}(2023){Guilera}, {Benitez-Llambay}, {Miller Bertolami}, \& {Pessah}}]{Guilera2023}
{Guilera}, O.~M., {Benitez-Llambay}, P., {Miller Bertolami}, M.~M., \& {Pessah}, M.~E. 2023, \bibinfo{title}{{Quantifying the Impact of the Dust Torque on the Migration of Low-mass Planets},} \apj, 953, 97, \dodoi{10.3847/1538-4357/acd2cb}

\bibitem[{O.~M. {Guilera} {et~al.}(2025){Guilera}, {Benitez-Llambay}, {Miller Bertolami}, \& {Pessah}}]{Guilera2025}
{Guilera}, O.~M., {Benitez-Llambay}, P., {Miller Bertolami}, M.~M., \& {Pessah}, M.~E. 2025, \bibinfo{title}{{Quantifying the Impact of the Dust Torque on the Migration of Low-mass Planets. II. The Role of Pebble Accretion in Planet Growth within a Global Planet Formation Model},} \apj, 986, 199, \dodoi{10.3847/1538-4357/add92a}

\bibitem[{O.~M. {Guilera} {et~al.}(2020){Guilera}, {S{\'a}ndor}, {Ronco}, {Venturini}, \& {Miller Bertolami}}]{Guilera2020}
{Guilera}, O.~M., {S{\'a}ndor}, Z., {Ronco}, M.~P., {Venturini}, J., \& {Miller Bertolami}, M.~M. 2020, \bibinfo{title}{{Giant planet formation at the pressure maxima of protoplanetary disks. II. A hybrid accretion scenario},} \aap, 642, A140, \dodoi{10.1051/0004-6361/202038458}

\bibitem[{Q. {Hou} {\&} C. {Yu}(2024){Hou} \& {Yu}}]{Hou2024}
{Hou}, Q., \& {Yu}, C. 2024, \bibinfo{title}{{Streaming Torque in Dust{\textendash}Gas Coupled Protoplanetary Disks},} \apj, 972, 152, \dodoi{10.3847/1538-4357/ad6a5c}

\bibitem[{Q. {Hou} {\&} C. {Yu}(2025){Hou} \& {Yu}}]{Hou2025}
{Hou}, Q., \& {Yu}, C. 2025, \bibinfo{title}{{Streaming Torque with Turbulent Diffusion},} \apj, 979, 185, \dodoi{10.3847/1538-4357/ada15a}

\bibitem[{H.-F. {Hsieh} {\&} M.-K. {Lin}(2020){Hsieh} \& {Lin}}]{Hsieh2020}
{Hsieh}, H.-F., \& {Lin}, M.-K. 2020, \bibinfo{title}{{Migrating low-mass planets in inviscid dusty protoplanetary discs},} \mnras, 497, 2425, \dodoi{10.1093/mnras/staa2115}

\bibitem[{J.~D. Hunter(2007)Hunter}]{Hunter:2007}
Hunter, J.~D. 2007, \bibinfo{title}{Matplotlib: A 2D graphics environment,} Computing in Science \& Engineering, 9, 90, \dodoi{10.1109/MCSE.2007.55}

\bibitem[{M.~A. {Jim{\'e}nez} {\&} F.~S. {Masset}(2017){Jim{\'e}nez} \& {Masset}}]{JimenezMasset2017}
{Jim{\'e}nez}, M.~A., \& {Masset}, F.~S. 2017, \bibinfo{title}{{Improved torque formula for low- and intermediate-mass planetary migration},} \mnras, 471, 4917, \dodoi{10.1093/mnras/stx1946}

\bibitem[{M. {Keppler} {et~al.}(2018){Keppler}, {Benisty}, {M{\"u}ller}, {Henning}, {van Boekel}, {Cantalloube}, {Ginski}, {van Holstein}, {Maire}, {Pohl}, {Samland}, {Avenhaus}, {Baudino}, {Boccaletti}, {de Boer}, {Bonnefoy}, {Chauvin}, {Desidera}, {Langlois}, {Lazzoni}, {Marleau}, {Mordasini}, {Pawellek}, {Stolker}, {Vigan}, {Zurlo}, {Birnstiel}, {Brandner}, {Feldt}, {Flock}, {Girard}, {Gratton}, {Hagelberg}, {Isella}, {Janson}, {Juhasz}, {Kemmer}, {Kral}, {Lagrange}, {Launhardt}, {Matter}, {M{\'e}nard}, {Milli}, {Molli{\`e}re}, {Olofsson}, {P{\'e}rez}, {Pinilla}, {Pinte}, {Quanz}, {Schmidt}, {Udry}, {Wahhaj}, {Williams}, {Buenzli}, {Cudel}, {Dominik}, {Galicher}, {Kasper}, {Lannier}, {Mesa}, {Mouillet}, {Peretti}, {Perrot}, {Salter}, {Sissa}, {Wildi}, {Abe}, {Antichi}, {Augereau}, {Baruffolo}, {Baudoz}, {Bazzon}, {Beuzit}, {Blanchard}, {Brems}, {Buey}, {De Caprio}, {Carbillet}, {Carle}, {Cascone}, {Cheetham}, {Claudi}, {Costille}, {Delboulb{\'e}}, {Dohlen}, {Fantinel}, {Feautrier}, {Fusco}, {Giro},
  {Gluck}, {Gry}, {Hubin}, {Hugot}, {Jaquet}, {Le Mignant}, {Llored}, {Madec}, {Magnard}, {Martinez}, {Maurel}, {Meyer}, {M{\"o}ller-Nilsson}, {Moulin}, {Mugnier}, {Orign{\'e}}, {Pavlov}, {Perret}, {Petit}, {Pragt}, {Puget}, {Rabou}, {Ramos}, {Rigal}, {Rochat}, {Roelfsema}, {Rousset}, {Roux}, {Salasnich}, {Sauvage}, {Sevin}, {Soenke}, {Stadler}, {Suarez}, {Turatto}, \& {Weber}}]{Keppler2018}
{Keppler}, M., {Benisty}, M., {M{\"u}ller}, A., {et~al.} 2018, \bibinfo{title}{{Discovery of a planetary-mass companion within the gap of the transition disk around PDS 70},} \aap, 617, A44, \dodoi{10.1051/0004-6361/201832957}

\bibitem[{M. {Lambrechts} {et~al.}(2019){Lambrechts}, {Morbidelli}, {Jacobson}, {Johansen}, {Bitsch}, {Izidoro}, \& {Raymond}}]{Lambrechts2019}
{Lambrechts}, M., {Morbidelli}, A., {Jacobson}, S.~A., {et~al.} 2019, \bibinfo{title}{{Formation of planetary systems by pebble accretion and migration. How the radial pebble flux determines a terrestrial-planet or super-Earth growth mode},} \aap, 627, A83, \dodoi{10.1051/0004-6361/201834229}

\bibitem[{T.~C.~H. {Lau} {et~al.}(2022){Lau}, {Dr{\k{a}}{\.z}kowska}, {Stammler}, {Birnstiel}, \& {Dullemond}}]{Lau2022}
{Lau}, T. C.~H., {Dr{\k{a}}{\.z}kowska}, J., {Stammler}, S.~M., {Birnstiel}, T., \& {Dullemond}, C.~P. 2022, \bibinfo{title}{{Rapid formation of massive planetary cores in a pressure bump},} \aap, 668, A170, \dodoi{10.1051/0004-6361/202244864}

\bibitem[{C. {Marois} {et~al.}(2008){Marois}, {Macintosh}, {Barman}, {Zuckerman}, {Song}, {Patience}, {Lafreni{\`e}re}, \& {Doyon}}]{Marois2008}
{Marois}, C., {Macintosh}, B., {Barman}, T., {et~al.} 2008, \bibinfo{title}{{Direct Imaging of Multiple Planets Orbiting the Star HR 8799},} Science, 322, 1348, \dodoi{10.1126/science.1166585}

\bibitem[{F.~S. {Masset}(2001){Masset}}]{Masset2001}
{Masset}, F.~S. 2001, \bibinfo{title}{{On the Co-orbital Corotation Torque in a Viscous Disk and Its Impact on Planetary Migration},} \apj, 558, 453, \dodoi{10.1086/322446}

\bibitem[{F.~S. {Masset}(2017){Masset}}]{Masset2017}
{Masset}, F.~S. 2017, \bibinfo{title}{{Coorbital thermal torques on low-mass protoplanets},} \mnras, 472, 4204, \dodoi{10.1093/mnras/stx2271}

\bibitem[{F.~S. Masset {et~al.}(2006)Masset, D’Angelo, \& Kley}]{Masset_2006}
Masset, F.~S., D’Angelo, G., \& Kley, W. 2006, \bibinfo{title}{On the Migration of Protogiant Solid Cores,} The Astrophysical Journal, 652, 730, \dodoi{10.1086/507515}

\bibitem[{F.~S. {Masset} {\&} J.~C.~B. {Papaloizou}(2003){Masset} \& {Papaloizou}}]{masset2003}
{Masset}, F.~S., \& {Papaloizou}, J.~C.~B. 2003, \bibinfo{title}{{Runaway Migration and the Formation of Hot Jupiters},} apj, 588, 494, \dodoi{10.1086/373892}

\bibitem[{A. {Morbidelli}(2020){Morbidelli}}]{Morbidelli2020}
{Morbidelli}, A. 2020, \bibinfo{title}{{Planet formation by pebble accretion in ringed disks},} \aap, 638, A1, \dodoi{10.1051/0004-6361/202037983}

\bibitem[{Y. {Nakagawa} {et~al.}(1986){Nakagawa}, {Sekiya}, \& {Hayashi}}]{Nakagawa1986}
{Nakagawa}, Y., {Sekiya}, M., \& {Hayashi}, C. 1986, \bibinfo{title}{{Settling and growth of dust particles in a laminar phase of a low-mass solar nebula},} \icarus, 67, 375, \dodoi{10.1016/0019-1035(86)90121-1}

\bibitem[{S. {Orcajo} {et~al.}(2025){Orcajo}, {Cieza}, {Guilera}, {P{\'e}rez}, {Rannou}, {Gonz{\'a}lez-Ruilova}, {Batalla-Falcon}, {Bhowmik}, {Chavan}, {Casassus}, {Dasgupta}, {Diaz}, {Gomez}, {Hales}, {Miley}, {Miller Bertolami}, {Nogueira}, {Ronco}, {Ruiz-Rodriguez}, {Sierra}, {Venturini}, {Weber}, {Williams}, \& {Zurlo}}]{Orcajo2025}
{Orcajo}, S., {Cieza}, L.~A., {Guilera}, O., {et~al.} 2025, \bibinfo{title}{{The Ophiuchus DIsk Survey Employing ALMA (ODISEA): A Unified Evolutionary Sequence of Planet-driven Substructures Explaining the Diversity of Disk Morphologies},} \apjl, 984, L57, \dodoi{10.3847/2041-8213/adcd58}

\bibitem[{S. {Paardekooper} {et~al.}(2023){Paardekooper}, {Dong}, {Duffell}, {Fung}, {Masset}, {Ogilvie}, \& {Tanaka}}]{Paardekooper2023}
{Paardekooper}, S., {Dong}, R., {Duffell}, P., {et~al.} 2023, \bibinfo{title}{{Planet-Disk Interactions and Orbital Evolution},} in Astronomical Society of the Pacific Conference Series, Vol. 534, Protostars and Planets VII, ed. S.~{Inutsuka}, Y.~{Aikawa}, T.~{Muto}, K.~{Tomida}, \& M.~{Tamura}, 685, \dodoi{10.48550/arXiv.2203.09595}

\bibitem[{S.-J. {Paardekooper} {et~al.}(2011){Paardekooper}, {Baruteau}, \& {Kley}}]{Paardekooper2011}
{Paardekooper}, S.-J., {Baruteau}, C., \& {Kley}, W. 2011, \bibinfo{title}{{A torque formula for non-isothermal Type I planetary migration - II. Effects of diffusion},} \mnras, 410, 293, \dodoi{10.1111/j.1365-2966.2010.17442.x}

\bibitem[{X.~S. {Ramos} {\&} P. {Ben{\'\i}tez-Llambay}(2026){Ramos} \& {Ben{\'\i}tez-Llambay}}]{Ramos2025}
{Ramos}, X.~S., \& {Ben{\'\i}tez-Llambay}, P. 2026, \bibinfo{title}{{A Comoving Framework for Planet Migration},} \apj, 998, 346, \dodoi{10.3847/1538-4357/ae36ab}

\bibitem[{Z. {Reg{\'a}ly}(2020){Reg{\'a}ly}}]{Regaly2020}
{Reg{\'a}ly}, Z. 2020, \bibinfo{title}{{Torques felt by solid accreting planets},} \mnras, 497, 5540, \dodoi{10.1093/mnras/staa2181}

\bibitem[{Z. {S{\'a}ndor} {et~al.}(2024{\natexlab{a}}){S{\'a}ndor}, {Guilera}, {Reg{\'a}ly}, \& {Lyra}}]{Sandor2024}
{S{\'a}ndor}, Z., {Guilera}, O.~M., {Reg{\'a}ly}, Z., \& {Lyra}, W. 2024{\natexlab{a}}, \bibinfo{title}{{Planetesimal and planet formation in transient dust traps},} \aap, 686, A78, \dodoi{10.1051/0004-6361/202347605}

\bibitem[{{}{Savvidou S.} {\&} {}{Bitsch B.}(2025){Savvidou S.} \& {Bitsch B.}}]{Savvidou2025}
{Savvidou S.}, \& {Bitsch B.} 2025, \bibinfo{title}{There is no disk mass budget problem of planet formation,} A\&A, 693, A302, \dodoi{10.1051/0004-6361/202449847}

\bibitem[{N. Shakura {\&} R. Sunyaev(1973)Shakura \& Sunyaev}]{Shakura1973}
Shakura, N., \& Sunyaev, R. 1973, \bibinfo{title}{Black holes in binary systems. Observational appearance,} Astronomy and Astrophysics, 24, 337

\bibitem[{H. {Tanaka} {et~al.}(2002){Tanaka}, {Takeuchi}, \& {Ward}}]{Tanaka2002}
{Tanaka}, H., {Takeuchi}, T., \& {Ward}, W.~R. 2002, \bibinfo{title}{{Three-Dimensional Interaction between a Planet and an Isothermal Gaseous Disk. I. Corotation and Lindblad Torques and Planet Migration},} \apj, 565, 1257, \dodoi{10.1086/324713}

\bibitem[{P. {Weber} {et~al.}(2019){Weber}, {P{\'e}rez}, {Ben{\'\i}tez-Llambay}, {Gressel}, {Casassus}, \& {Krapp}}]{Weber2019}
{Weber}, P., {P{\'e}rez}, S., {Ben{\'\i}tez-Llambay}, P., {et~al.} 2019, \bibinfo{title}{{Predicting the Observational Signature of Migrating Neptune-sized Planets in Low-viscosity Disks},} \apj, 884, 178, \dodoi{10.3847/1538-4357/ab412f}

\end{thebibliography}
\end{document}